\documentclass[pra,onecolumn,showpacs]{revtex4-1}
\usepackage{makeidx}
\usepackage{amssymb}
\usepackage{amsmath}
\usepackage{graphicx}
\usepackage{subfigure}
\usepackage{color}
\usepackage{verbatim}
\usepackage{titletoc}
\usepackage[hidelinks]{hyperref}

\begin{document}

\title{ Ring modes supported by concentrated cubic nonlinearity}
\author{Elad Shamriz and Boris A. Malomed}

\begin{abstract}
We consider the one-dimensional Schr\"{o}dinger equation on a ring, with the
cubic term, of either self-attractive or repulsive sign, confined to a
narrow segment. This setting can be realized in optics and Bose-Einstein
condensates. For the nonlinearity coefficient represented by the
delta-function, all stationary states are obtained in an exact analytical
form. The states with positive chemical potentials are found in alternating
bands for the cases of the self-repulsion and attraction, while solutions
with negative chemical potentials exist only in the latter case. These
results provide a possibility to obtain exact solutions for bandgap states
in the nonlinear system. Approximating the delta-function by a narrow
Gaussian, stability of the stationary modes is addressed through numerical
computation of eigenvalues for small perturbations, and verified by
simulations of the perturbed evolution. For positive chemical potentials,
the stability is investigated in three lowest bands. In the case of the
self-attraction, each band contains a stable subband, the transition to
instability occurring with the increase of the total norm. As a result,
multi-peak states may be stable in higher bands. In the case of the
self-repulsion, a single-peak ground state is stable in the first band,
while the two higher ones are populated by weakly unstable two- and
four-peak excited states. In the case of the self-attraction and negative
chemical potentials, single-peak modes feature instability which transforms
them into persistently oscillating states.
\end{abstract}

\maketitle

\address{Department of Physical Electronics, School of Electrical
Engineering, Faculty of Engineering, and the Center for Light-Matter
Interaction, Tel Aviv University, P.O.B. 39040, Ramat Aviv, Tel
Aviv, Israel}

\section{Introduction and the model}

The use of spatially modulated nonlinearities makes it possible to greatly
expand varieties of solitons supported by competition of local
nonlinearities and linear diffraction or dispersion \cite{FA,PGK,RMP}. The
simplest example of such settings is represented by a continuous linear
medium into which a cubic self-focusing nonlinearity is embedded in a narrow
region, that may be approximated by the coefficient in front of the cubic
term taken as the delta-function, $\delta (x)$. This approximation leads to
the following limit form of the nonlinear Schr\"{o}dinger equation (NLSE)
for wave function $\psi \left( x,t\right) $ [alias the Gross-Pitaevskii
equation, in terms of atomic Bose-Einstein condensates (BECs) \cite{GPE}],%
\begin{equation}
i\psi _{t}+\frac{1}{2}\psi _{xx}+\varepsilon \delta (x)|\psi |^{2}\psi =0,
\label{delta(x)}
\end{equation}%
written in the scaled form, with strength $\varepsilon >0$ of the
self-focusing. This model was introduced in Ref. \cite{Mark}, where the
scattering problem for a plane wave hitting the localized nonlinearity was
considered, and localized modulational instability of the solution to the
scattering problem was discovered. Note that $\varepsilon $ may be fixed as
an arbitrary positive value by means of obvious rescaling of wave function $%
\psi $.

In the application to optics, with time $t$ in Eq. (\ref{delta(x)}) replaced
by the propagation distance, $z$ \cite{KA}, a narrow nonlinearity-bearing
stripe embedded in a planar nonlinear waveguide may be created by implanting
nonlinear dopants into the host linear medium \cite{Kip}. In BEC, a similar
effect can be achieved by the locally induced Feshbach resonance, controlled
by a tightly focused laser beam, as suggested by the techniques demonstrated
in Refs. \cite{Painting} and \cite{FR1}-\cite{FR3}.

NLSE in the form of Eq. (\ref{delta(x)}) gives rise to an obvious family of
pinned modes (solitons), parameterized by chemical potential $\mu <0$,
\begin{equation}
\psi =\left( -2\mu \right) ^{1/4}\exp \left( -i\mu t-\sqrt{-2\mu }|x|\right)
\label{sol1}
\end{equation}%
(in terms of the NLSE for the propagation of light in planar waveguides, $%
-\mu $ is the propagation constant). The norm of the solitons (alias the
integral power of the optical beam),%
\begin{equation}
N=\int_{-\infty }^{+\infty }\left\vert \psi (x)\right\vert ^{2}dx,  \label{N}
\end{equation}%
is degenerate for family (\ref{sol1}), taking a single value which does not
depend on $\mu $, $N\equiv 1$. The norm degeneracy is a characteristic
feature of families of \textit{Townes solitons} (TS) \cite{Townes}, which
exist in models featuring the critical collapse driven by self-attractive
nonlinearities \cite{Berge,Fibich}. Accordingly, this family formally seems
neutrally stable in terms of the well-known Vakhitov-Kolokolov (VK)
criterion, $dN/d\mu <0$, which often plays the role of the necessary
stability criterion for self-trapped states maintained by attractive
nonlinearities \cite{VK,Berge,Fibich}. However, in reality solitons of the
TS type are subject to nonlinear (subexponentially growing) instability,
which destroys them \cite{Berge,Fibich}. Indeed, the soliton family (\ref%
{sol1}) is completely unstable, in the framework of Eq. (\ref{delta(x)})
\cite{Nir}. It is relevant to mention that, while the original TS family is
represented by axially symmetric solutions of the two-dimensional NLSE with
the cubic nonlinearity \cite{Townes}, a similar family is known in one
dimension too, in the form of the NLSE with the quintic self-focusing \cite%
{1DTownes}. In fact, the soliton family (\ref{sol1}) may be considered as an
alternative example of the TS family in one dimension.

Because below we also consider the model with the repulsive localized
nonlinearity, $\varepsilon <0$ in Eq. (\ref{NLS}), it is relevant to mention
that, in cases when the repulsive nonlinearity may support self-trapped
modes (e.g., gap solitons, in the presence of a spatially periodic potential
\cite{gap1}-\cite{gap3}), their necessary stability condition may amount\ to
the \textit{anti-VK} criterion, $dN/d\mu >0$ \cite{HS}. Note that the
localized repulsive nonlinear term can be efficiently used as a splitter in
the design of soliton-based matter-wave interferometers \cite{HS2}.

The instability of the TS family in the framework of the fundamental model (%
\ref{delta(x)}) makes it necessary to look for physically relevant
modifications of the model, which may stabilize solitons maintained by the
strongly localized nonlinearity; actually, this implies the necessity to
lift the TS norm degeneracy \cite{BenLi}. One possibility is to add a
spatially periodic linear potential to Eq. (\ref{delta(x)}) \cite{Nir}, and
another is to consider a set of two localized nonlinearities, both
represented by the delta-function \cite{Thaw,Shnir}. In the latter case,
only self-trapped modes which keep the symmetry with respect to the pair of
delta-functions, are stable, while replacement of the ideal delta-functions
by a regularized approximation, see Eq. (\ref{Gauss}) below, creates
stability regions for antisymmetric and asymmetric modes too (they exist as
completely unstable exact solutions in the case of the pair of ideal
delta-functions).

Another possibility is to consider Eq. (\ref{delta(x)}) on a ring, i.e., to
rewrite it in the scaled form, with $x$ replaced by the angular coordinate
defined in the interval of $-\pi <\theta <+\pi $:

\begin{equation}
i\psi _{t}+\frac{1}{2}\psi _{\theta \theta }+\varepsilon \delta (\theta
)|\psi ^{2}|\psi =0,  \label{NLS}
\end{equation}%
subject to the periodic boundary conditions (b.c.):

\begin{eqnarray}
\psi (\theta &=&-\pi ,t)=\psi (\theta =+\pi ,t),  \notag \\
\psi _{\theta }(\theta &=&-\pi ,t)=\psi _{\theta }(\theta =+\pi ,t)
\label{psi_BC}
\end{eqnarray}%
The energy (Hamiltonian) of the system is
\begin{equation}
H=\frac{1}{2}\left[ \int_{-\pi }^{+\pi }\left\vert \psi _{\theta
}\right\vert ^{2}d\theta -\varepsilon \left\vert \psi \left( \theta
=0\right) \right\vert ^{4}\right] .  \label{E}
\end{equation}

As mentioned above, the strength of the localized nonlinearity in Eq. (\ref%
{NLS}), $\varepsilon $, may be fixed by the rescaling of wave functions $%
\psi $, if the analysis admits variation of its norm (\ref{N}). For the of
presentation of results in a compact form, in the case of self-attraction, $%
\varepsilon >0$, it is convenient to fix $\varepsilon =+3$, and in the case
of repulsion a convenient choice is $\varepsilon =-1$, which is adopted
below, although these normalizations do not have any specific significance.

The ring-shaped setting can be implemented in diverse physical settings,
including toroidal traps for BEC, which were proposed theoretically \cite%
{BECring-theory} and realized in many experiments \cite{BECring1}-\cite%
{weak-link3}. In optics, the ring model implies guided light propagation
along cylindrical surfaces, which has also been reported in various forms,
such as concentric multilayer \textit{omniguiding fibers} with a hollow core
\cite{omni1}-\cite{omni3}, multilayer fibers which provide Bragg confinement
in the radial direction \cite{Bragg1}-\cite{Koby}, concentric structures
built in photorefractive materials \cite{Hoq,Photorefr-radial}, and laser
sources in the form of VCSELs \cite{VCSEL1}-\cite{VCSEL3}. In addition to
optics, the guided transmission of plasmonic waves in narrow cylindrical
layers has been realized too \cite{plasmonic-fiber,plasmonic-fiber2}. These
settings make it possible to impart topological characteristics to photonic
modes, the vorticity being the simplest one. The topological structure may
protect various modes against perturbations in photonics \cite{Topol-phot}
(as well as in BEC\ \cite{protectedBEC} and acoustics \cite{acoustics}), an
important recently introduced example being provided by surface modes in
diverse realizations of photonic topological insulators \cite{PhotInsulator1}%
-\cite{Segev4}

While the ring models are often introduced as linear ones, they may readily
include nonlinearity, which makes it possible to predict solitons localized
in the azimuthal direction \cite{rotary,rotary2,Bakhtiyor}. In particular,
states supported by a periodic modulation of the local nonlinearity in the
rings were studied in Ref. \cite{Kuzmiak}. Further, the cubic nonlinearity
in the ring may be localized in a narrow segment, as implied by Eq. (\ref%
{NLS}). In particular, similar settings in BEC loaded in toroidal traps have
been created with narrow \textquotedblleft weak links" embedded in the
respective ring-shaped configurations \cite{weak-link1,weak-link2,weak-link3}%
.

A model similar to one based on Eq. (\ref{NLS}), but with a pair of
self-attractive ($\varepsilon >0$) delta-functional nonlinear spots set at
diametrically opposite points, was introduced in Ref. \cite{HPu}. The
analysis of the model, which was limited solely to states with the negative
chemical potential [i.e., $\psi (x)$ composed solely of hyperbolic
functions] had led to a conclusion that, similar to what was reported for
the infinite system in Ref. \cite{Thaw}, only modes symmetric with respect
to the pair of two nonlinearity spots may be stable in the case of ideal
delta-functions, while the replacement of them by regularized approximations
gives rise to stable asymmetric modes too. Although the ring-shaped system
with the single delta-functional nonlinearity seems simpler and, in a sense,
more fundamental, it was not studied \ before, being the subject of the
current work, for both signs of the chemical potential, $\mu \gtrless 0$,
and both signs of $\varepsilon $ in Eq. (\ref{NLS}). As mentioned above, $%
\varepsilon <0$ implies the localized repulsive cubic term, which is also
possible in BEC and optics, but was not considered in Ref. \cite{HPu}.

In the model elaborated in the present work, states with $\mu <0$ (they
exist only for $\varepsilon >0$) are qualitatively similar to those reported
for the pair of ideal delta-functions in Ref. \cite{HPu}. The most essential
results are reported for $\mu >0$, with both signs of $\varepsilon $. This
case was not addressed in Ref. \cite{HPu}, as it is difficult to obtain
respective analytical solutions, composed of trigonometric functions, for
the pair of delta-functions in the ring. The results produced here for $\mu
>0$ provide direct insight into the structure of \emph{nonlinear bandgap
modes} in the form of exact analytical solutions, which is not available in
other models, to the best of our knowledge. In particular, we report exact
solutions for both single-peak ground states and multi-peak excited modes,
while their stability is studied with the help of numerical methods. This is
done by means of a solution of the eigenvalue problem \cite{Yang} for the
linearized Bogoliubov - de Gennes (BdG) equations for small perturbations
\cite{GPE}. The predictions produced by the calculation of the BdG
eigenvalues are validated by direct simulations of the perturbed evolution,
while formal predictions of the VK criterion are not completely correct in
the present model.

The numerical results are obtained with the ideal delta-function replaced by
a narrow Gaussian with small width $\xi $, as specified below by Eq. (\ref%
{Gauss}). In this connection, it is relevant to discuss how realistic the
use of the delta-function is for modeling physical settings. In BEC, the
width of the nonlinear layer, induced by the optically-controlled Feshbach
resonance, cannot be smaller than the respective wavelength, $\sim 1$ $%
\mathrm{\mu }$m. On the other hand, the ring-shaped quasi-one-dimensional
trap can be created with diameter $\simeq 3$ mm \cite{BECring2}, hence, in
the scaled form, the regularization parameter in Eq. (\ref{Gauss}) is
bounded by condition $\xi \gtrsim 10^{-4}$. In optics, the technique of the
thermal indiffusion makes it possible to create highly doped stripes of
width $\simeq 4~\mathrm{\mu }$m \cite{Kip}, while the the VCSEL-like
structure can be made with diameter $\simeq 300$ $\mathrm{\mu }$m \cite%
{VCSEL}, thus corresponding to $\xi \simeq 0.004$. In the numerical
calculations, we chiefly use $\xi =0.01$ and $0.005$ (in some cases), which
are relevant values, in terms of these estimates. Such values of $\xi $
produce numerical results which are extremely close to the analytical ones
obtained in the model with the ideal delta-function. Furthermore, additional
numerical considerations demonstrate that the results remain practically the
same (in particular, as concerns the stability), at least, up to $\xi $ $%
\simeq 0.2$.

In fact, in the case of $\varepsilon <0$ there is no essential constraint on
the size of $\xi $, while in the case of self-focusing, $\varepsilon >0$,
there is a constraint imposed by the condition of the modulational stability
of the field in the nonlinear layer of width $\xi $. A simple estimate
demonstrates that the modulational instability does not set in if amplitude $%
\phi _{0}$ of the field\ at $\theta =0$ satisfies constraint%
\begin{equation}
\phi _{0}^{2}<2/\left( \varepsilon \xi \right) .  \label{no-MI}
\end{equation}%
All the results presented below meet this condition.

The rest of the paper is organized as follows. In Section II, analytical
solutions for stationary modes are displayed, for the ideal delta-function
with both signs of $\varepsilon $ in Eq. (\ref{NLS}), and both signs of $\mu
$. Numerical results for stationary solutions and their stability are
reported in several parts of Section III, and the paper is concluded by
Section IV.

\section{Analytical solutions}

Stationary solutions to Eqs. (\ref{NLS}) and (\ref{psi_BC}) with chemical
potential $\mu $ are looked for as $\psi (\theta )=e^{-i\mu t}\phi (\theta )$%
, where real function $\phi (\theta )$ satisfies equation

\begin{equation}
\mu \phi +\frac{1}{2}\phi ^{\prime \prime }+\varepsilon \delta (\theta )\phi
^{3}=0,  \label{stationary_eq}
\end{equation}%
(with $\phi ^{\prime }\equiv d\phi /d\theta $) at $-\pi <\theta <+\pi $,
supplemented by the periodic b.c. which are set at $\theta =\pm \pi $, as
per Eq. (\ref{psi_BC}):

\begin{equation}
\phi (-\pi )=\phi (+\pi ),\phi ^{\prime }(-\pi )=\phi ^{\prime }(+\pi )~.
\label{phi_BC}
\end{equation}%
Equation (\ref{stationary_eq}) implies that one should actually solve the
linear equation,%
\begin{equation}
\phi ^{\prime \prime }+2\mu \phi =0,  \label{linear}
\end{equation}%
separately at $-\pi <\theta <0$ and $0<\theta <+\pi $, subjecting them to
b.c. (\ref{phi_BC}), and to the condition for the jump of the first
derivative at $\theta =0$, which follows from the integration of Eq. (\ref%
{phi_BC}) in an infinitesimal vicinity of $\theta =0$:

\begin{equation}
\phi ^{\prime }|_{\theta =+0}-\phi ^{\prime }|_{\theta =-0}=-2\varepsilon
\phi ^{3}|_{\theta =0},  \label{der_jump}
\end{equation}%
while $\phi (\theta )$ is continuous at $\theta =0$.

\subsection{Solutions for $\protect\mu >0$}

As said above, most interesting are solutions for positive values of the
chemical potential, and both signs of $\varepsilon $, as similar exact
results were not reported in previous studies. A relevant solution to linear
equation (\ref{linear}), satisfying b.c. (\ref{phi_BC}), is%
\begin{equation}
\phi (\theta )=\phi _{0}\cos \left( \sqrt{2\mu }\left( \pi -|\theta |\right)
\right) ,  \label{ansatz}
\end{equation}%
where real amplitude $\phi _{0}$ is found from the substitution of
expression (\ref{ansatz}) in b.c. (\ref{der_jump}) at $\theta =0$:

\begin{equation}
\phi _{0}^{2}=-\frac{\sqrt{2\mu }\sin \left( \pi \sqrt{2\mu }\right) }{%
\varepsilon \cos ^{3}\left( \pi \sqrt{2\mu }\right) }
\label{wave_amplitude_first_form}
\end{equation}%
The calculation of the integrals in Eqs. (\ref{N}) and (\ref{E}) for this
solution yields its norm and energy:%
\begin{gather}
N=\phi _{0}^{2}\left[ \pi +\frac{1}{2\sqrt{2\mu }}\sin \left( 2\sqrt{2\mu }%
\pi \right) \right] ,  \label{N mu>0} \\
H=\mu \phi _{0}^{2}\left[ \pi -\frac{1}{2\sqrt{2\mu }}\sin \left( 2\sqrt{%
2\mu }\pi \right) \right] -\frac{\varepsilon }{2}\phi _{0}^{4}\cos
^{4}\left( \sqrt{2\mu }\pi \right) ,  \label{E1}
\end{gather}%
The energy can be obtained in the form of $E=E(N)$ by eliminating $\mu $
from Eqs. (\ref{N mu>0}) and (\ref{E1}).

An obvious condition for the existence of this solution is $\phi _{0}^{2}>0$%
. First, at $\varepsilon >0$ (the self-attractive nonlinearity), it follows
from this condition and Eq. (\ref{wave_amplitude_first_form}) that $\mu $
must satisfy inequality $\tan \left( \pi \sqrt{2\mu }\right) <0$, hence the
solutions for $\varepsilon >0$ exist in the following \textit{bands}
(intervals of values of the chemical potential):

\begin{equation}
\frac{1}{2}\left( \frac{1}{2}+n\right) ^{2}<\mu <\frac{1}{2}\left(
1+n\right) ^{2},~n=0,1,2,...~.  \label{mu_for_pos_epsi}
\end{equation}%
In the opposite case of the repulsive nonlinearity, $\varepsilon <0$, Eq. (%
\ref{wave_amplitude_first_form}) yields $\phi _{0}^{2}>0$ in a set of bands
alternating with those given by Eq. (\ref{mu_for_pos_epsi}):
\begin{equation}
\frac{1}{2}n^{2}<\mu <\frac{1}{2}\left( n+\frac{1}{2}\right)
^{2},~n=0,1,2,...\,.  \label{mu_for_neg_epsi}
\end{equation}

\begin{figure}[tbp]
\subfigure[]{\includegraphics[width=2.6in]{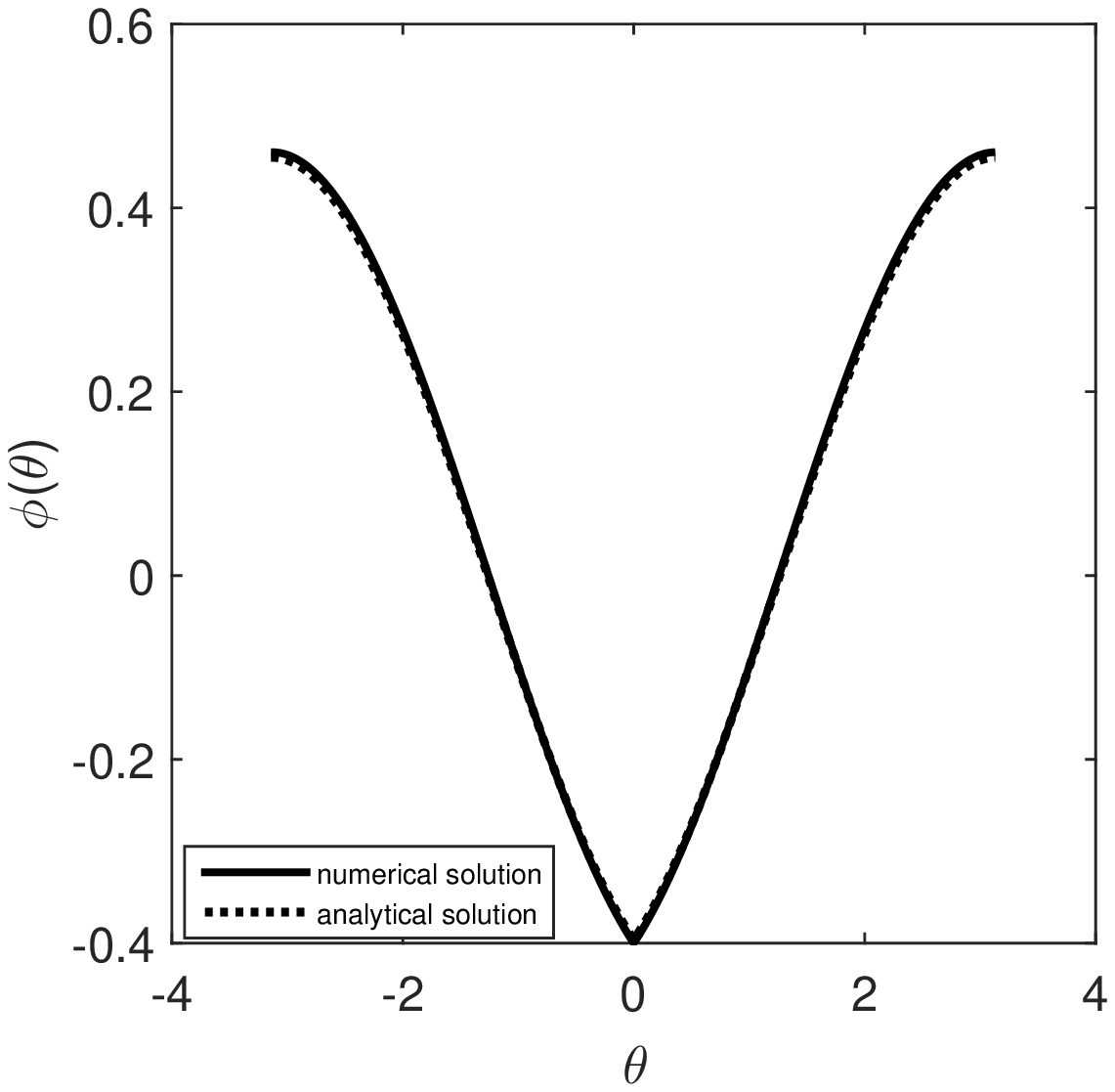}} \subfigure[]{%
\includegraphics[width=2.6in]{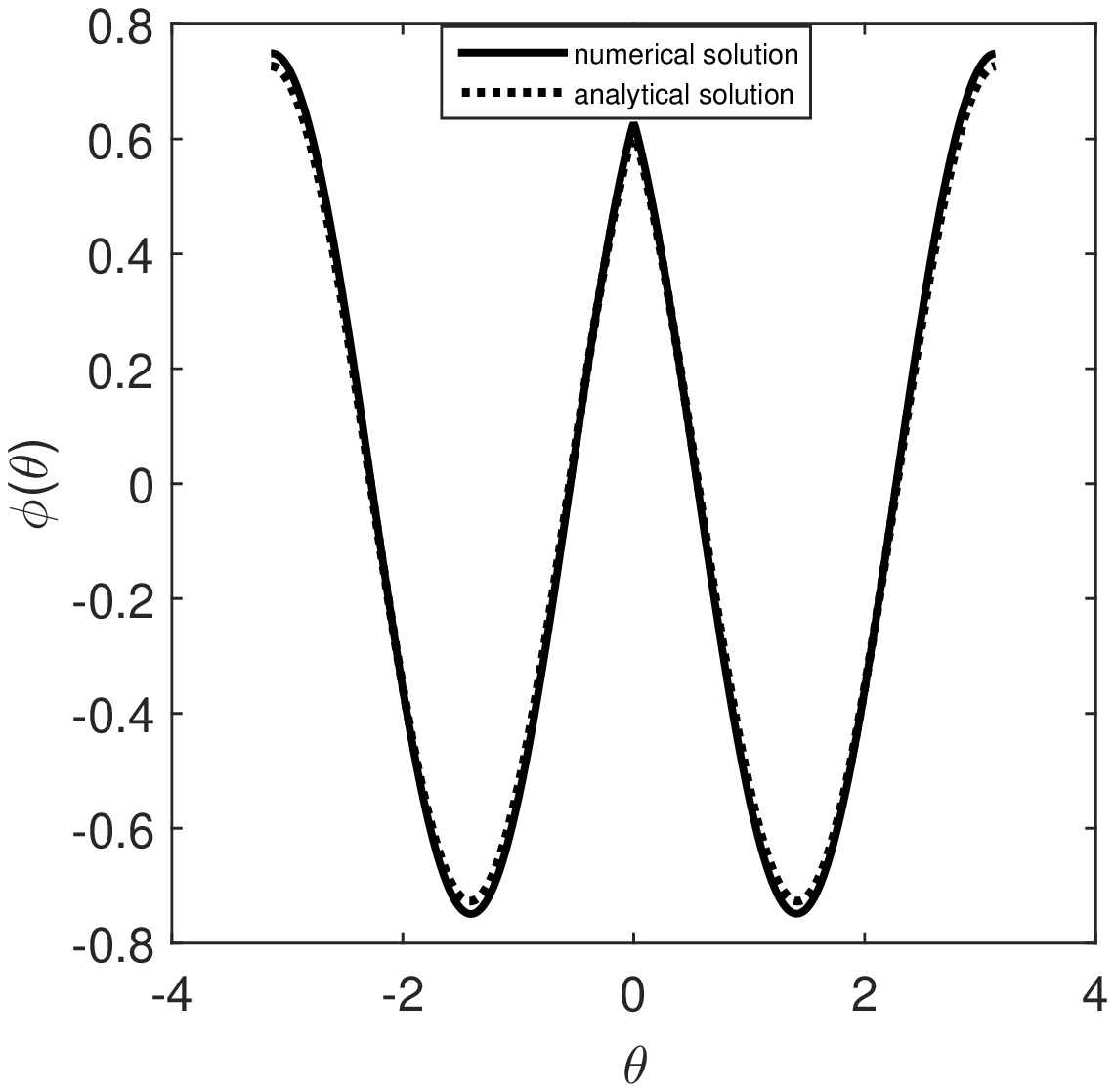}} \subfigure[]{%
\includegraphics[width=2.6in]{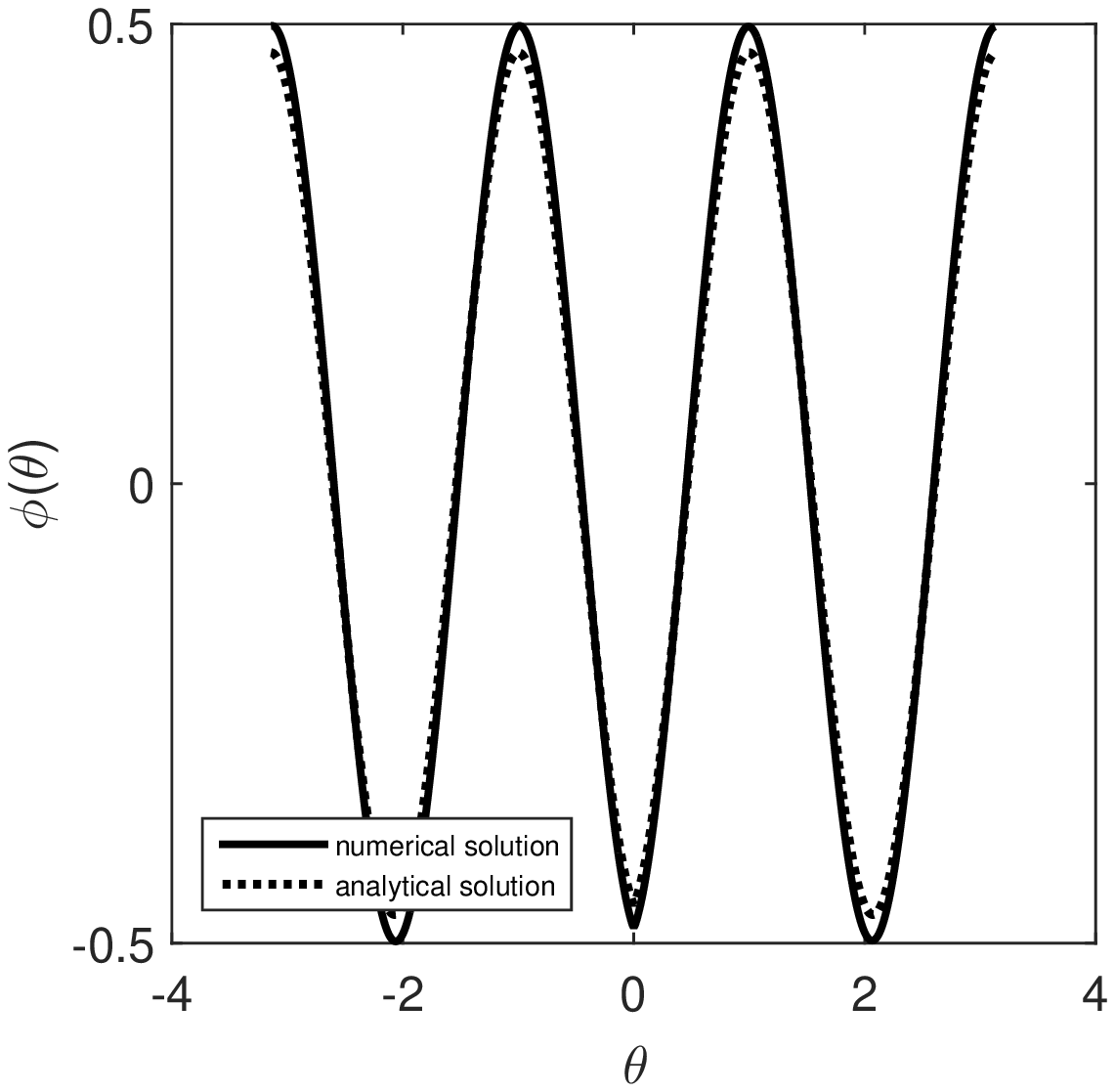}} \subfigure[]{%
\includegraphics[width=2.6in]{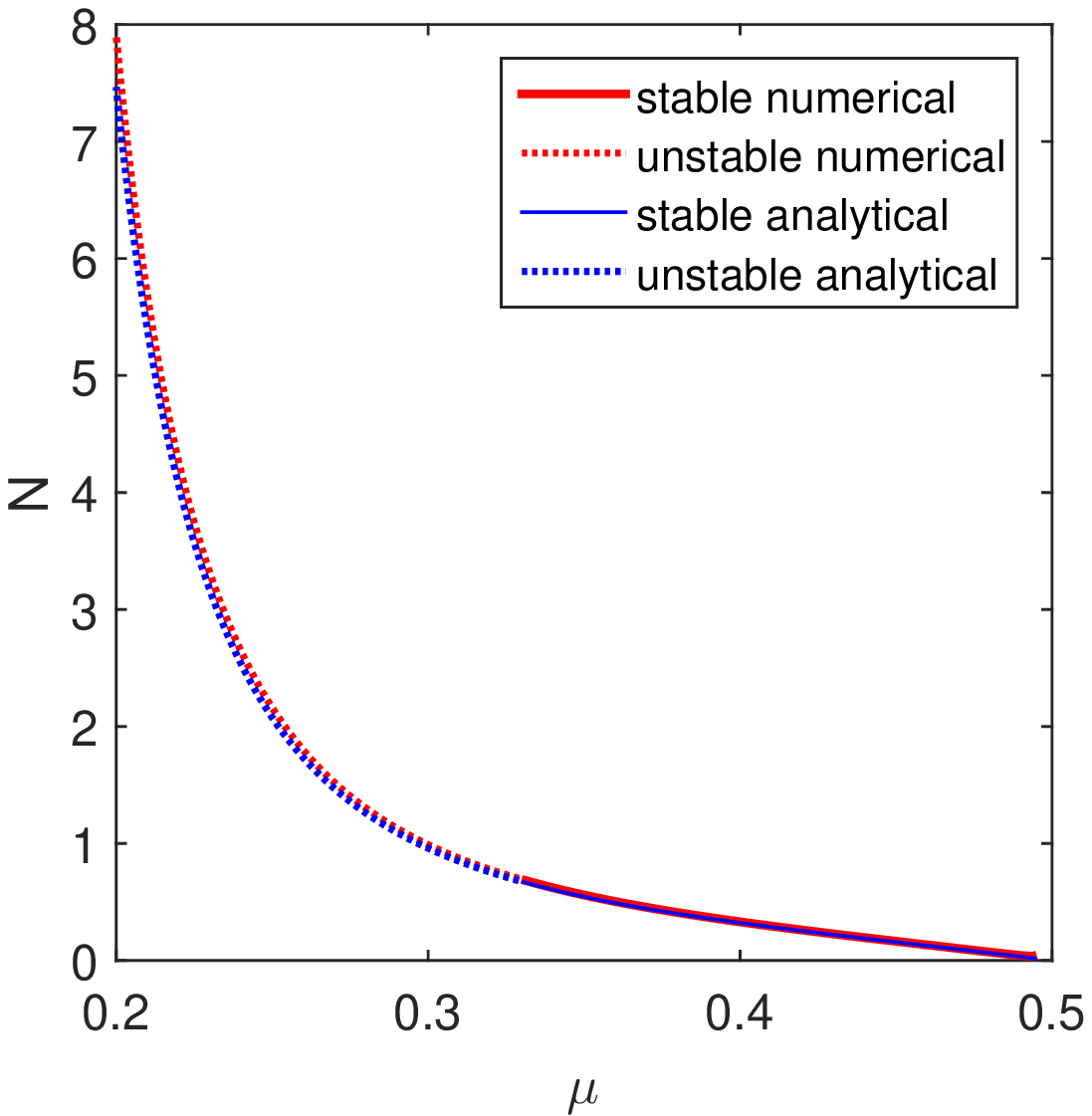}} \subfigure[]{%
\includegraphics[width=2.6in]{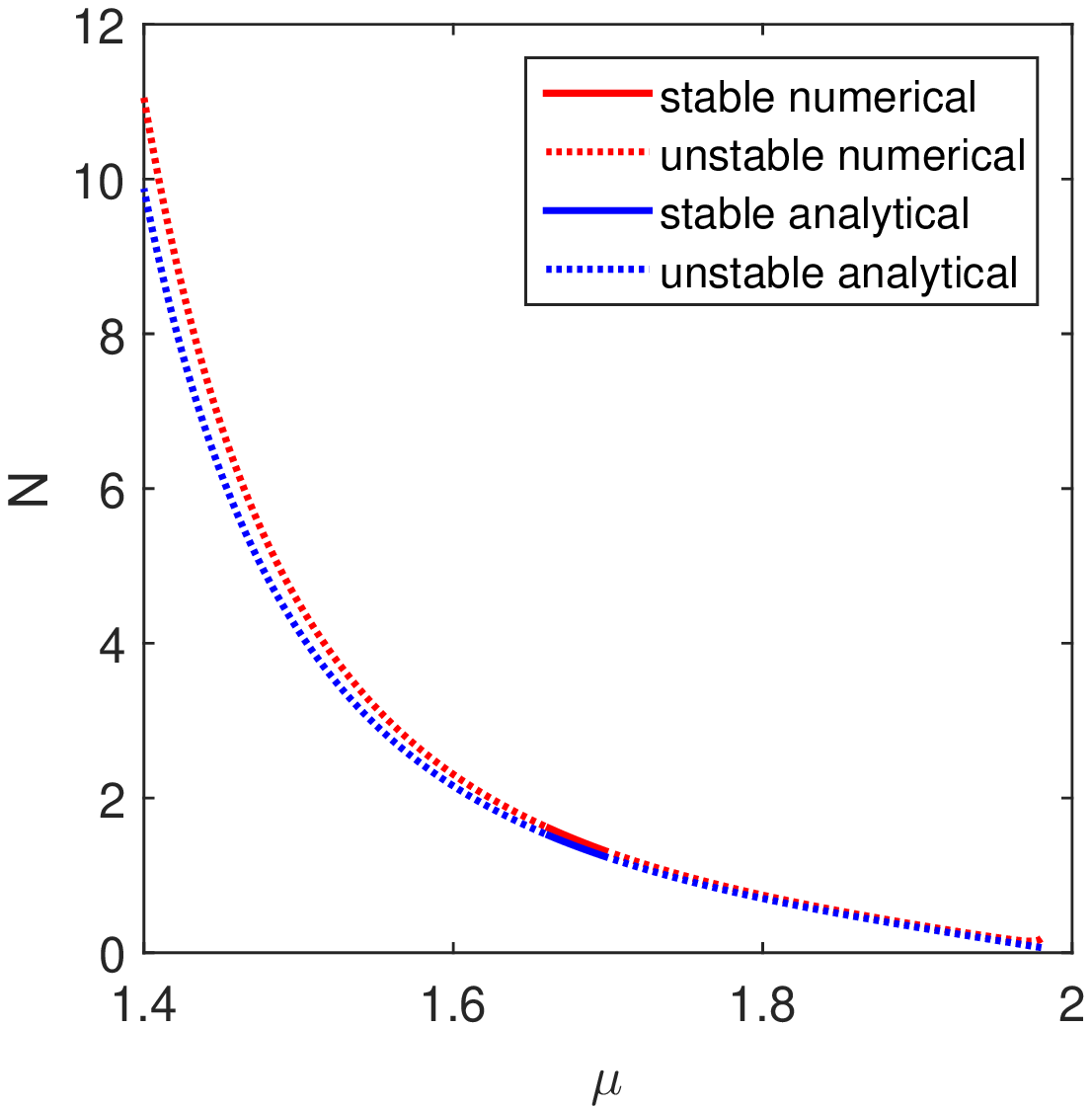}} \subfigure[]{%
\includegraphics[width=2.6in]{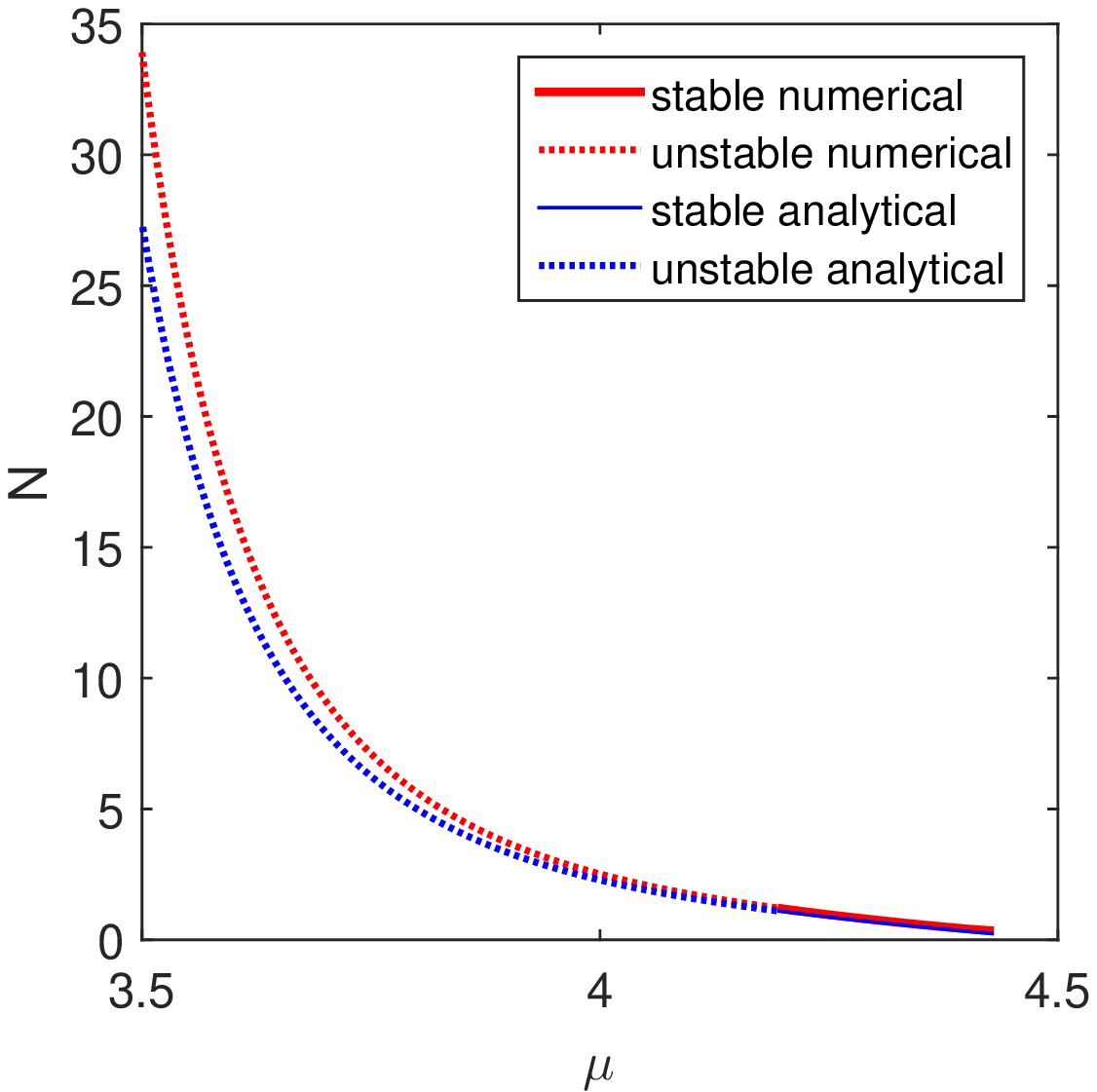}}
\caption{(a-c): Typical examples of stable analytical modes, given by Eqs. (%
\protect\ref{ansatz}) and (\protect\ref{wave_amplitude_first_form}) for $%
\protect\varepsilon =3$, and their counterparts, produced by the numerical
solution of Eq. (\protect\ref{stationary_eq}) using the regularized
delta-function (\protect\ref{Gauss}), with $\protect\xi =0.01$. In panels
(a), (b), and (c), the stationary solutions are displayed, severally, with
the chemical potentials and norms $\left( \protect\mu =0.35,N_{\mathrm{%
numerical}}=0.55,N_{\mathrm{analytical}}=0.54\right) $; $\left( \protect\mu %
=1.66,N_{\mathrm{numerical}}=1.57,N_{\mathrm{analytical}}=1.53\right) $; and
$\left( \protect\mu =4.3,N_{\mathrm{numerical}}=0.69,N_{\mathrm{analytical}%
}=0.67\right) $, which places them in the first, second, and third bands, as
defined by Eq. (\protect\ref{mu_for_pos_epsi}) with $n=0$, $1$, and $2$,
respectively. (d-f): $N(\protect\mu )$ curves for the analytical solutions,
calculated as per Eqs. (\protect\ref{N mu>0}) and (\protect\ref%
{wave_amplitude_first_form}), and their numerically generated counterparts,
in the first, second, and third bands, respectively. The full bands
correspond to intervals $0.125<\protect\mu <0.5$ ($n=0$), $1.125<\protect\mu %
<2$ ($n=1$), and $3.125<\protect\mu <4$ ($n=2$). Portions of the $N(\protect%
\mu )$ curves, representing stable and unstable subfamilies of the
stationary solutions, are designated as indicated in the notation boxes.
Definitely unstable segments of the $N(\protect\mu )$ curves, which
correspond to extremely large values of $N$, are cut off by panel frames.}
\label{Fig1}
\end{figure}

\begin{figure}[tbp]
\subfigure[]{\includegraphics[width=2.6in]{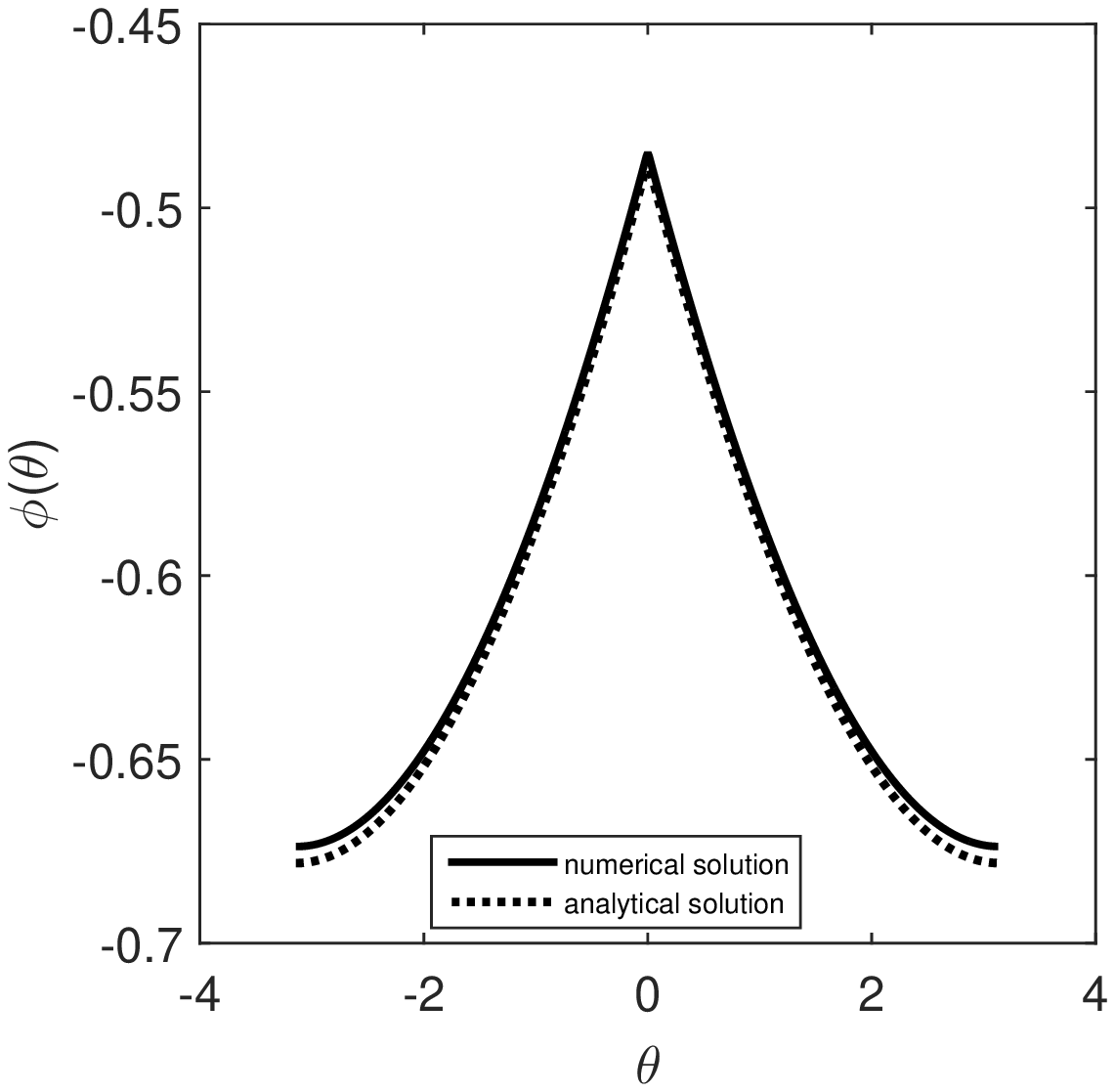}} \subfigure[]{%
\includegraphics[width=2.6in]{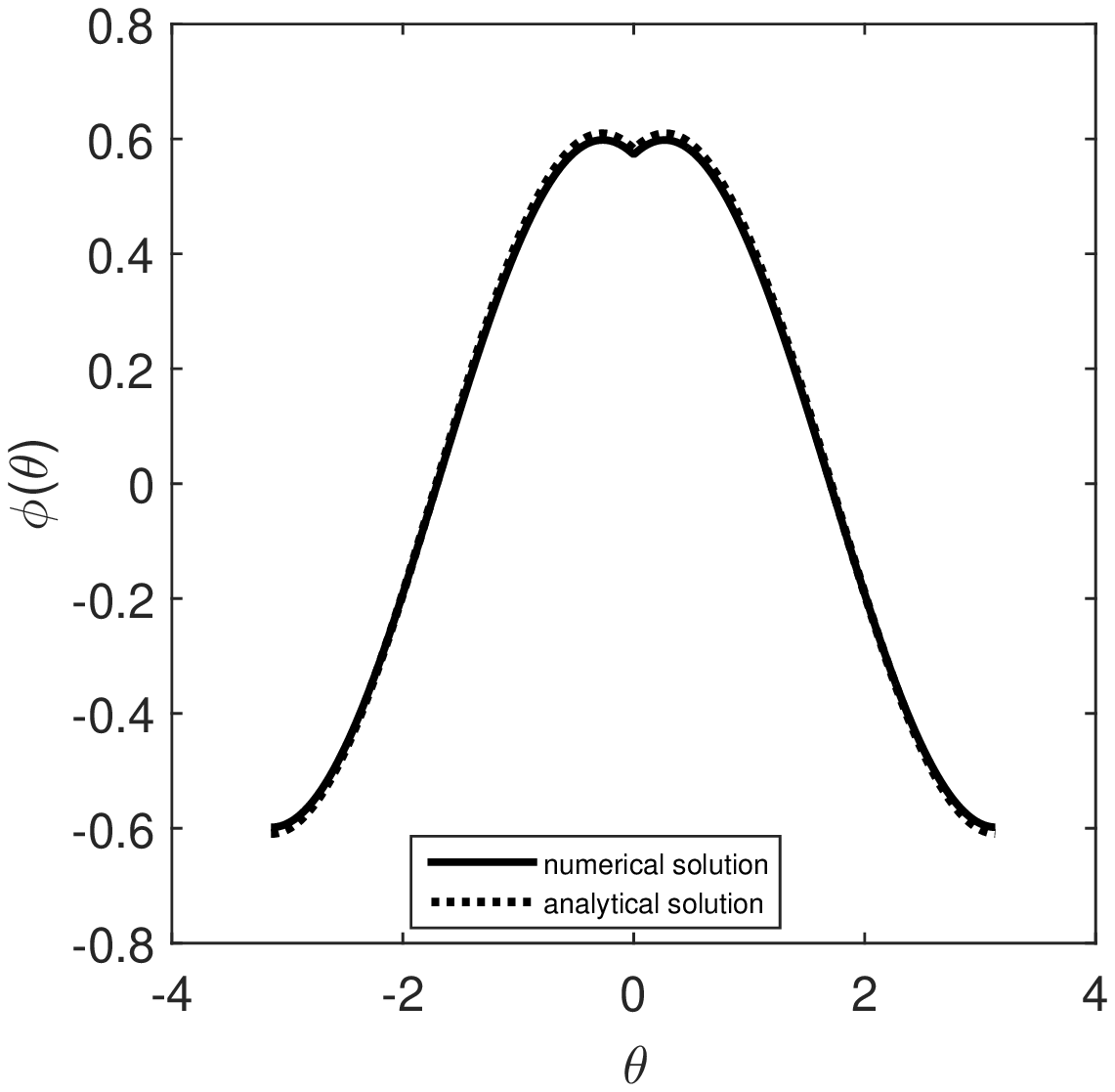}} \subfigure[]{%
\includegraphics[width=2.6in]{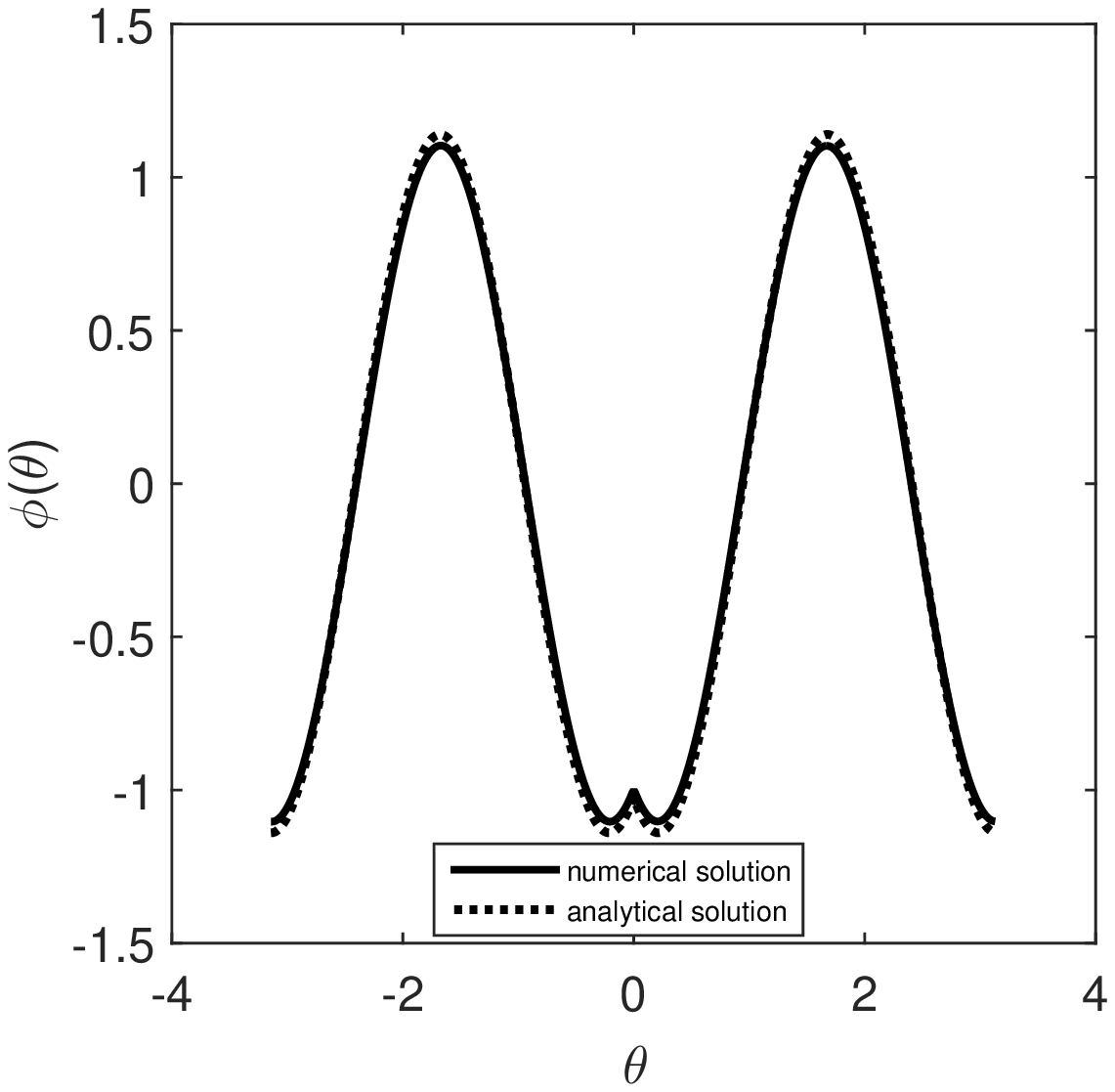}} \subfigure[]{%
\includegraphics[width=2.6in]{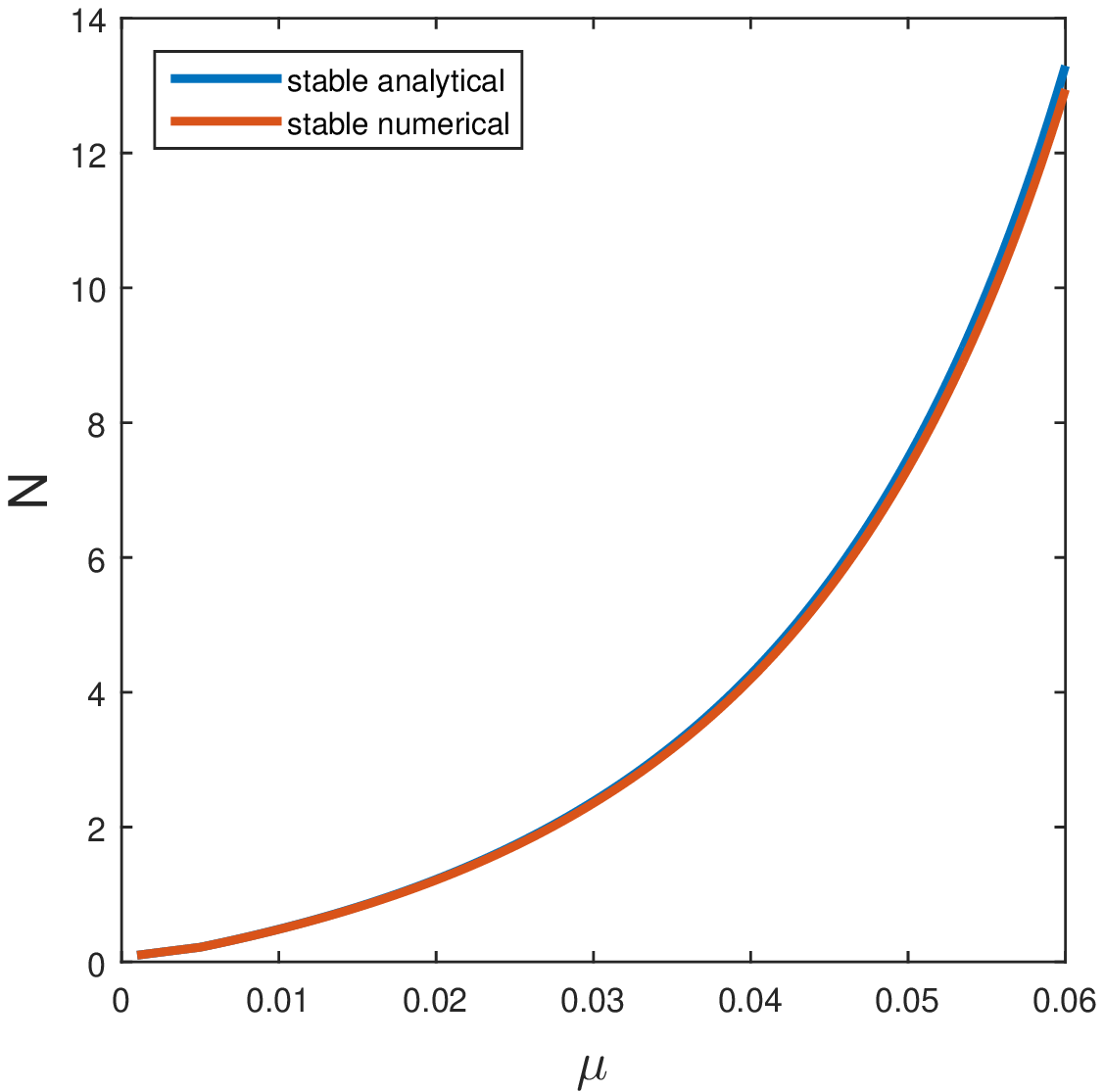}}
\caption{(a-c): Typical examples of analytical modes, given by Eqs. (\protect
\ref{ansatz}) and (\protect\ref{wave_amplitude_first_form}) with $\protect%
\varepsilon =-1$, and their counterparts, produced by the numerical solution
with $\protect\xi =0.01$ in Eq. (\protect\ref{Gauss}). In panels (a), (b),
and (c) the chemical potential and norm take values $\left( \protect\mu %
=0.03,N_{\mathrm{numerical}}=2.35,N_{\mathrm{analytical}}=2.38\right) $, $%
\left( \protect\mu =0.6,N_{\mathrm{numerical}}=1.24,N_{\mathrm{analytical}%
}=1.26\right) $, and $\left( \protect\mu =2.3,N_{\mathrm{numerical}}=4.20,N_{%
\mathrm{analytical}}=4.32\right) $, which places them in the first, second,
and third bands, as defined by Eq. (\protect\ref{mu_for_neg_epsi}) with $n=0$%
, $1$, and $2$, respectively. The mode shown in (a) is a stable ground
state, while ones in (b) and (c) are weakly unstable excited states. (d) The
$N(\protect\mu )$ curve for the analytical solution and its counterpart,
produced by the numerical solution of Eq. (\protect\ref{N mu>0}) with $%
\protect\xi =0.01$, in the first band, which covers the interval of $0<%
\protect\mu <0.125$. This branch, representing the ground state, is
completely stable. The branches of the excited states, populating the second
and third bands, are not displayed, as they are subject to weak instability.}
\label{Fig2}
\end{figure}

Typical examples of the exact solutions, and the respective dependences $%
N(\mu )$, juxtaposed with their numerical counterparts, are displayed, for $%
\varepsilon >0$ and $\varepsilon <0$, in Figs. \ref{Fig1} and \ref{Fig2},
respectively (as mentioned above, the cases of $\varepsilon >0$ and $%
\varepsilon <0$ are represented, severally, by $\varepsilon =+3$ and $%
\varepsilon =-1$). The stability/instability indicated in the figures is
identified as per results of the analysis presented below. In addition to
that, the analytical expression (\ref{E1}) and its numerically generated
counterpart demonstrate that, quite naturally, the energy of stationary
states populating different bands is much higher in higher-order bands, for
the same values of $N$ (not shown here in detail).

In terms of the local density, $\phi _{0}^{2}(x)$, the modes shown in Fig. %
\ref{Fig1} for $\varepsilon >0$ in the $n$-th band feature $2(1+n)$ peaks
[i.e., $2$, $4$, and $6$ peaks in the first, second, and third bands,
respectively, see Eq. (\ref{mu_for_pos_epsi})]. On the other hand, the modes
displayed for $\varepsilon <0$ in Fig. \ref{Fig4}, have, essentially, $2n$
peaks [i.e., $1$, $2$, and $4$ peaks in the first, second, and third bands,
respectively, as per Eq. (\ref{mu_for_neg_epsi})], if the shallow splitting
of the peak at $x=0$ is not counted for $n=1$ and $2$. In particular, the
stationary mode in the first band, with the single density peak at $\theta
=\pm \pi $, which is displayed in Fig. \ref{Fig2}(a) (and is always stable,
as shown below), may be identified as the ground state of the system with $%
\varepsilon <0$, as its energy, for given $N$, is always lowest, in
comparison with the states found in the second and third bands. On the other
hand, the multi-peak solutions, which populate the second, third (and
higher-order) bands at $\varepsilon <0$, may be interpreted as excited
states, always being weakly unstable, as shown below too. For $\varepsilon
>0 $, the identification of a ground state makes it necessary to consider
solutions with $\mu <0$, which are addressed in the following subsection
(recall that states with $\mu <0$ do not exist for $\varepsilon <0$).

It is worthy to note that all $N(\mu )$ curves displayed in Fig. \ref{Fig1}
meet the above-mentioned VK criterion, $dN/d\mu <0$, thus having a chance to
be (partly or entirely) stable. Further, in Fig. \ref{Fig2}(d) the $N(\mu )$
dependence for $\varepsilon <0$ satisfies the anti-VK criterion \cite{HS}, $%
dN/d\mu <0$, thus upholding the stability of the soliton family.

\subsection{Solutions for $\protect\mu <0$}

In the infinite domain, negative values of the chemical potential correspond
to pinned solitons (\ref{sol1}), which, as said above, are completely
unstable solutions. In the ring-shaped system, the relevant solution to Eq. (%
\ref{linear}), subject to b.c. (\ref{phi_BC}), is%
\begin{equation}
\phi (\theta )=\phi _{0}\cosh \left( \sqrt{-2\mu }\left( \pi -|\theta
|\right) \right) ,  \label{cosh}
\end{equation}%
\begin{equation}
\phi _{0}^{2}=\frac{\sqrt{-2\mu }\sinh \left( \sqrt{-2\mu }\pi \right) }{%
\varepsilon \cosh ^{3}\left( \sqrt{-2\mu }\pi \right) },  \label{phi0-cosh}
\end{equation}%
cf. Eqs. (\ref{ansatz}) and (\ref{wave_amplitude_first_form}). As follows
from Eq. (\ref{phi0-cosh}), this solution exists, for $\mu <0$, solely in
the case of the attractive nonlinearity, $\varepsilon >0$. The norm and
energy of the solution for $\mu <0$, given by Eqs. (\ref{cosh}) and \ref%
{phi0-cosh}), are
\begin{equation}
N=\phi _{0}^{2}\left[ \pi +\frac{1}{2\sqrt{-2\mu }}\sinh \left( 2\sqrt{-2\mu
}\pi \right) \right] ,  \label{N mu<0}
\end{equation}%
\begin{equation}
H=\left( -\mu \right) \phi _{0}^{2}\left[ \frac{1}{2\sqrt{-2\mu }}\sinh
\left( 2\sqrt{-2\mu }\pi \right) -\pi \right] -\frac{\varepsilon }{2}\phi
_{0}^{4}\cosh ^{4}\left( \sqrt{-2\mu }\pi \right) .  \label{E2}
\end{equation}

\begin{figure}[tbp]
\subfigure[]{\includegraphics[width=2.6in]{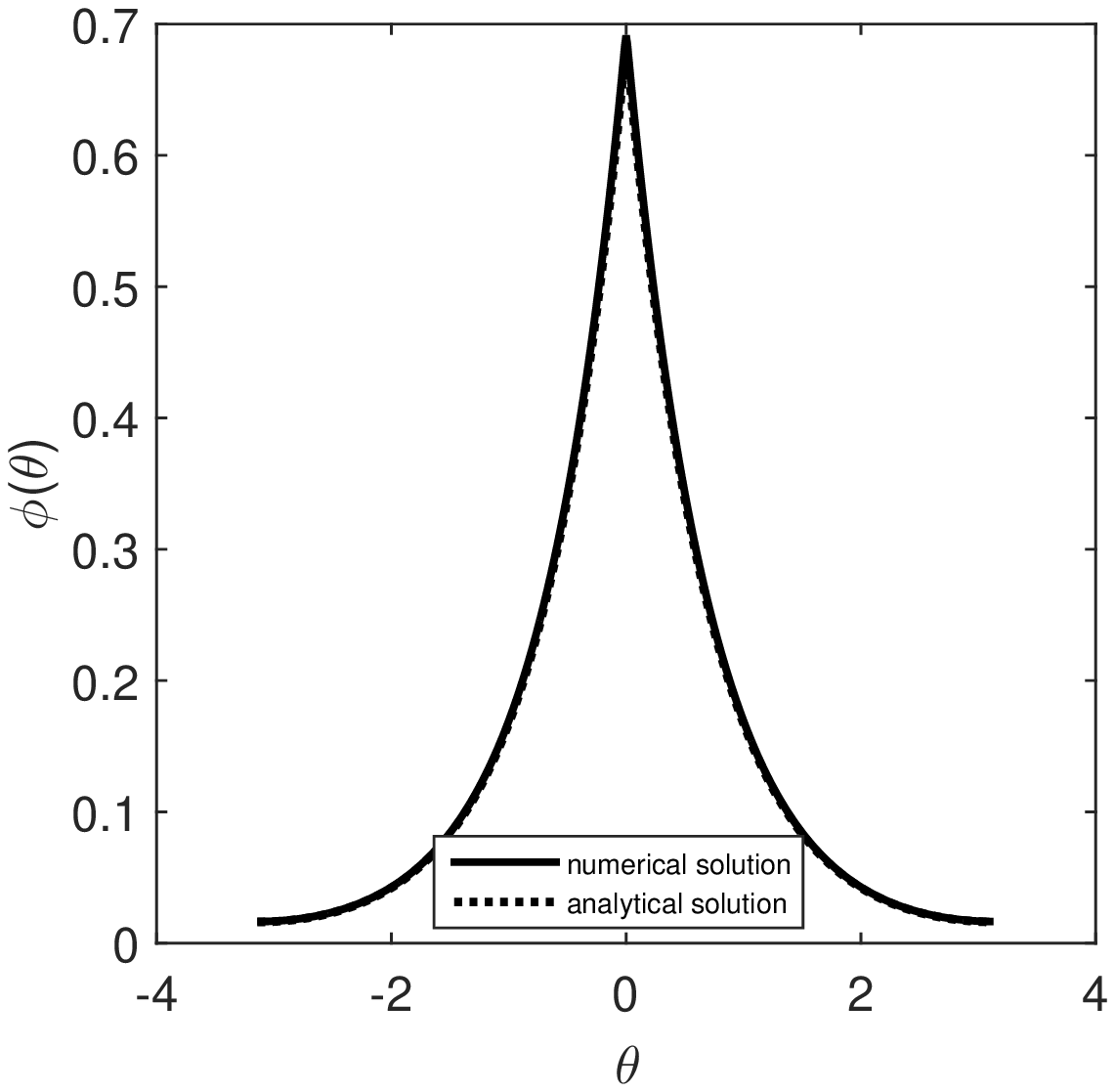}} \subfigure[]{%
\includegraphics[width=2.6in]{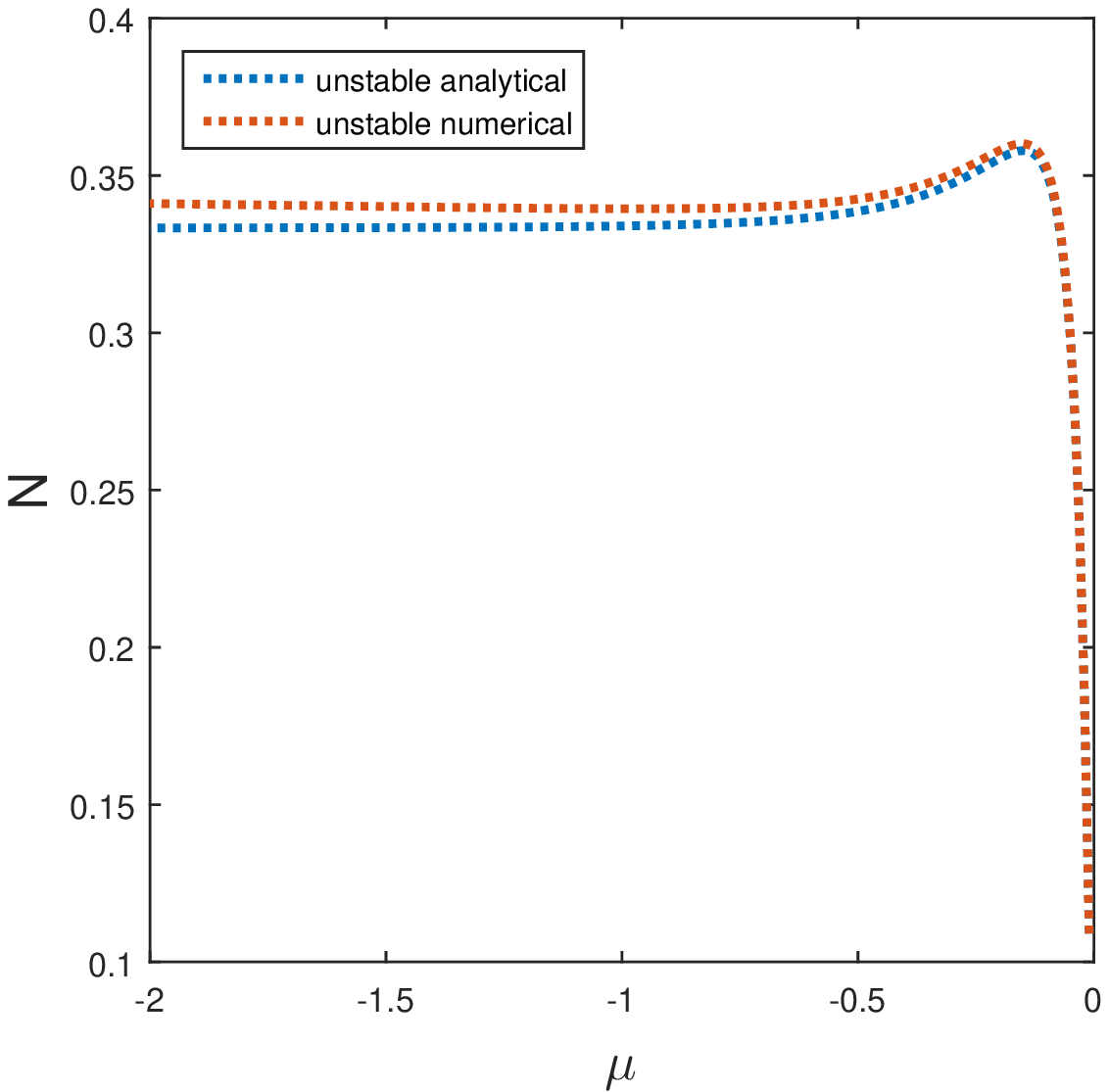}} \subfigure[]{%
\includegraphics[width=2.6in]{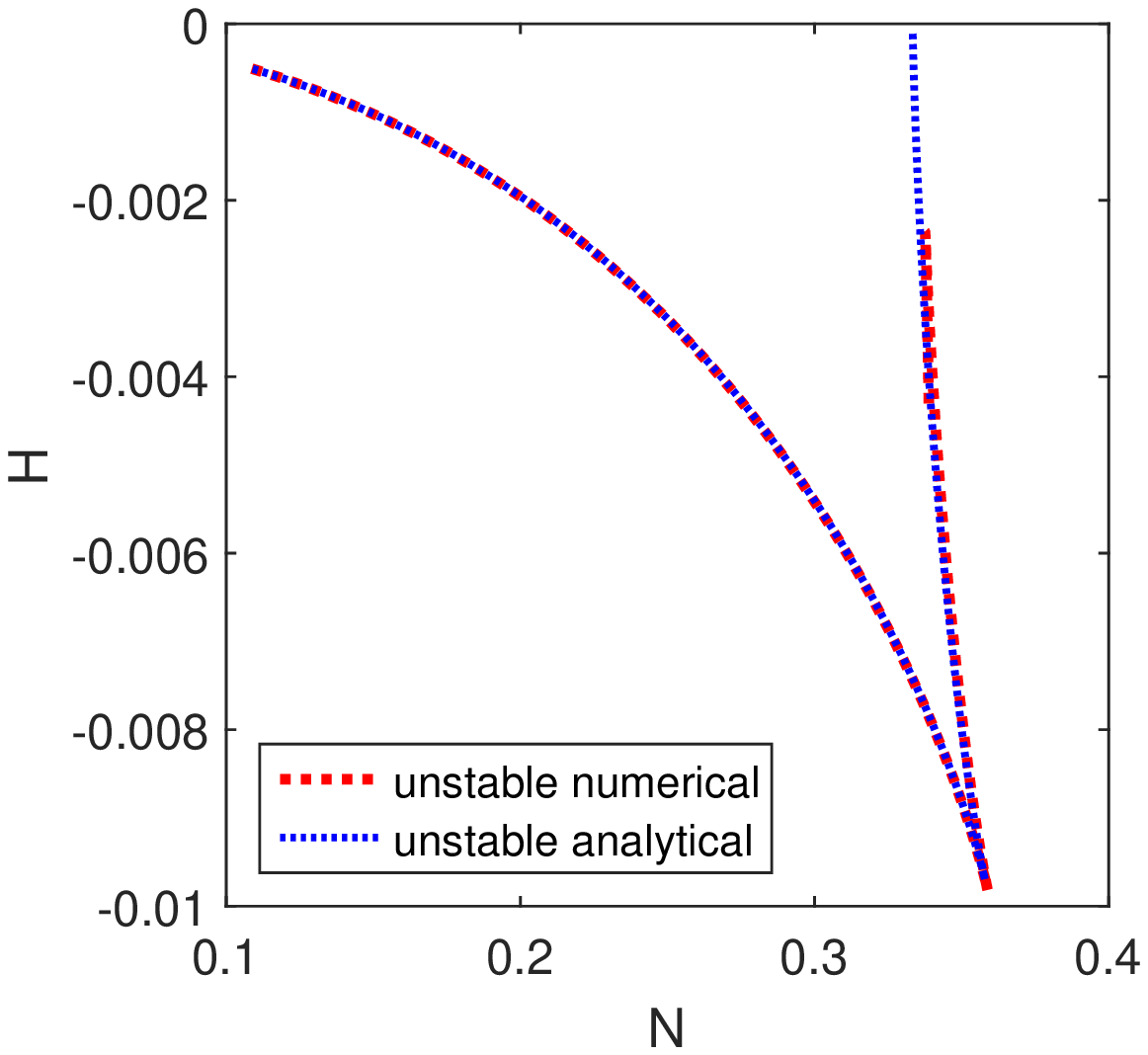}}
\caption{(a) A typical example of an [unstable, see Fig. \protect\ref{Fig4}%
(d) below] stationary mode existing at $\protect\mu <0$, as produced by the
analytical solution given by Eqs. (\protect\ref{cosh}) and (\protect\ref%
{phi0-cosh}), juxtaposed with the corresponding numerical solution of Eq. (%
\protect\ref{stationary_eq}), for $\protect\mu =-1$, $\protect\varepsilon =3$
and $\protect\xi =0.01$. The respective norms are $N_{\mathrm{numerical}%
}=0.342,N_{\mathrm{analytical}}=0.334$. (b,c) $N(\protect\mu )$ and $H(N)$
curves for the analytical solutions, produced by Eqs. (\protect\ref{N mu<0})
and (\protect\ref{E2}), and their numerically generated counterpart, for $%
\protect\mu <0$ and $\protect\varepsilon =3$. Unlike all other plots, the
numerical ones in this panel were generated with $\protect\xi =0.005$, to
achieve sufficient accuracy.}
\label{Fig3}
\end{figure}

It is relevant to note that the solution for $\mu <0$, given by Eqs. (\ref%
{cosh}), (\ref{phi0-cosh}) and (\ref{N mu<0}), (\ref{E2}), can be obtained
from the above one for $\mu >0$, based on Eqs. (\ref{ansatz}), (\ref%
{phi0-cosh}), and (\ref{N mu>0}), (\ref{E1}), as an analytical continuation
from $\mu >0$ to $\mu <0$, according to the straightforward relations:
\begin{equation}
\sqrt{2\mu }=i\sqrt{-2\mu },~\sin \left( \sqrt{2\mu }\pi \right) =i\sinh
\left( \sqrt{-2\mu }\pi \right) ,~\cos \left( \sqrt{2\mu }\pi \right) =\cosh
\left( \sqrt{-2\mu }\pi \right) .
\end{equation}%
\ A typical example of the mode with $\mu <0$, and the respective $N(\mu )$
and $H(N)$ dependences are displayed, along with their numerically found
counterparts, in Fig. \ref{Fig3}. Note that, unlike the solutions found
above at $\mu >0$, whose norm may take indefinitely large values, diverging
at the left edge of each band, as per Eqs. (\ref{N mu>0}) and (\ref%
{wave_amplitude_first_form}), straightforward analysis of Eqs. (\ref{N mu<0}%
) and (\ref{phi0-cosh}) reveals, as seen in Fig. \ref{Fig3}(b), that at $\mu
<0$ the norm is bounded from above by
\begin{equation}
N<N_{\max }\approx \allowbreak 1.\,\allowbreak 074/\varepsilon .
\label{Nmax}
\end{equation}%
This largest value of the norm is attained at
\begin{equation}
\mu _{\max }\approx -0.150.  \label{mumax}
\end{equation}%
Further, the $H(N)$ dependence at $\mu <0$ features two branches in Fig. \ref%
{Fig3}(c) in accordance with the fact that, in a narrow interval of norms, $%
N(\mu =-\infty )\equiv 1/\varepsilon <N<N_{\max }$ (i.e., $0.333<N<0.358$
for $\varepsilon =3$), two different values of $\mu $ correspond to given $N$%
. The solitons at $\mu <0$ always features a single density peak, like in
Fig. \ref{Fig3}(a).

The stability of these solutions is identified in the following section, by
means of numerical methods. Actually, they are always subject to a
(relatively weak) instability, even if the respective $N(\mu )$ dependence,
as seen in Fig. \ref{Fig3}(b), satisfies the VK criterion, $dN/d\mu <0$, in
an interval of
\begin{equation}
0<-\mu <-\mu _{\max }\approx 0.150.  \label{no-VK}
\end{equation}

Formally, the stationary solutions found at $\mu <0$ realize the ground
state of the system with $\varepsilon >0$, as seen from the comparison of
their negative energy, displayed in Fig. \ref{Fig3}(c), with the positive
energy of the solutions found, for the same values of $\varepsilon $ and $N$%
, at $\mu >0$. However, the instability of these stationary solutions
suggests that the role of the ground state may be picked up by robust
breathers which spontaneously develop from the unstable stationary states
with $\mu <0$, see Fig. \ref{Fig6} below.

\section{Numerical solutions}

\subsection{Stationary solutions}

In the numerical solution of the evolution and stationary equations (\ref%
{NLS}) and (\ref{stationary_eq}), the delta-function was approximated by the
standard regularized expression,
\begin{equation}
{\tilde{\delta}}(\theta )=\frac{1}{\sqrt{\pi }\xi }\exp \left( -\frac{\theta
^{2}}{\xi ^{2}}\right) ,  \label{Gauss}
\end{equation}%
with sufficiently small $\xi $. Stationary equation (\ref{stationary_eq})
with the regularized delta-function was numerically solved by means of the
Newton-Raphson method, which is a root-finding algorithm that uses a
truncated Taylor expansion to find a zero of a given function in a vicinity
of an expected zero point \cite{Yang}. Time-dependent solutions to Eq. (\ref%
{NLS}) were then simulated by means of the split-step fast-Fourier-transform
algorithm, which is well known to be appropriate for the NLSE \cite{Yang}.
The conservation of the integral norm and energy was monitored in all the
dynamical simulations.

\subsection{The linear-stability analysis: the Bogoliubov-de Gennes (BdG)
equations}

To analyze the stability of stationary solutions of Eq. (\ref{NLS}) against
small perturbations, perturbed solutions were taken in the usual form \cite%
{GPE,Yang}:%
\begin{equation}
\psi \left( \theta ,t\right) =e^{-i\mu t}\left\{ \phi (\theta )+\eta \left[
e^{-i\lambda t}u(\theta )+e^{i{\lambda }^{\ast }t}v^{\ast }(\theta )\right]
\right\} ,  \label{perturbation_1st_form}
\end{equation}%
where $\eta $ is an infinitesimal amplitude of the perturbation, $u(\theta )$
and $v(\theta )$ represent its eigenmode, and $\lambda $ is the
corresponding (generally, complex) perturbation eigenfrequency, the
stability condition being $\mathrm{Im}\{\lambda \}=0$ for all $\lambda $
(the asterisk stands for the complex conjugate). The substitution of this
expression in Eq. (\ref{NLS}) and linearization (i.e., the derivation of the
BdG equations for the small perturbations) leads, after straightforward
manipulations, to the eigenvalue problem for $\lambda $, written in the
matrix form:

\begin{equation}
\begin{pmatrix}
\hat{L} & -\varepsilon \delta (\theta )\phi ^{2}(\theta ) \\
\varepsilon \delta (\theta )\phi ^{2}(\theta ) & -\hat{L}%
\end{pmatrix}%
\begin{pmatrix}
u \\
v%
\end{pmatrix}%
=\lambda
\begin{pmatrix}
u \\
v%
\end{pmatrix}%
,  \label{eigen}
\end{equation}%
where~we define operator $\hat{L}\equiv -\mu -\frac{1}{2}d^{2}/d\theta
^{2}-2\varepsilon \delta (\theta )|\phi |^{2}$, and the solution for $%
\left\{ u(\theta ),v(\theta )\right\} $ must satisfy the same b.c. (\ref%
{psi_BC}) as above.

The numerical solution of Eq. (\ref{eigen}), with the delta-function
approximated as per Eq. (\ref{Gauss}), produces, for each stationary
solution, a spectrum of eigenfrequencies $\lambda $. The analysis of the
numerical data leads to conclusions about the stability of the modes with $%
\varepsilon >0$ (the self-attractive nonlinearity) and $\mu >0$, which are
displayed in Table 1. It identifies stable and unstable subbands in each of
the three lowest existence bands defined by Eq. (\ref{mu_for_pos_epsi}).

\begin{tabular}{|l|l|l|}
\hline
$\mathrm{first~band~}(n=0)$ & $\mathrm{second~band~}(n=1)$ & $\mathrm{%
third~band~}(n=2)$ \\ \hline
$0.125<\mu <0.5$ & $1.125<\mu <2$ & $3.125<\mu <4.5$ \\ \hline
$\mathrm{stability~subband}$ & $\mathrm{stability~subband}$ & $\mathrm{%
stability~subbands}$ \\ \hline
$0.337<\mu <0.5$ & $1.66<\mu <1.70$ & $4.194<\mu <4.5$ \\ \hline
\end{tabular}

\noindent {\small Table 1: Three lowest bands, corresponding to }${\small n=0%
}${\small , }${\small 1}${\small , and }${\small 2}${\small \ in Eq. (\ref%
{mu_for_pos_epsi})}$,${\small \ in which the exact solutions, given by Eqs. (%
\ref{ansatz}) and (\ref{wave_amplitude_first_form}), exists for }${\small %
\mu >0}${\small \ and }${\small \varepsilon >0}${\small , and subbands in
which they are stable, according to values of the perturbation
eigenfrequencies produced by the numerical solution of the BdG equations for
}${\small \varepsilon =3}$ {\small and regularization parameter }${\small %
\xi =0.01}$ {\small in Eq. (\ref{Gauss}).} \bigskip

\begin{figure}[tbp]
\subfigure[]{\includegraphics[width=2.6in]{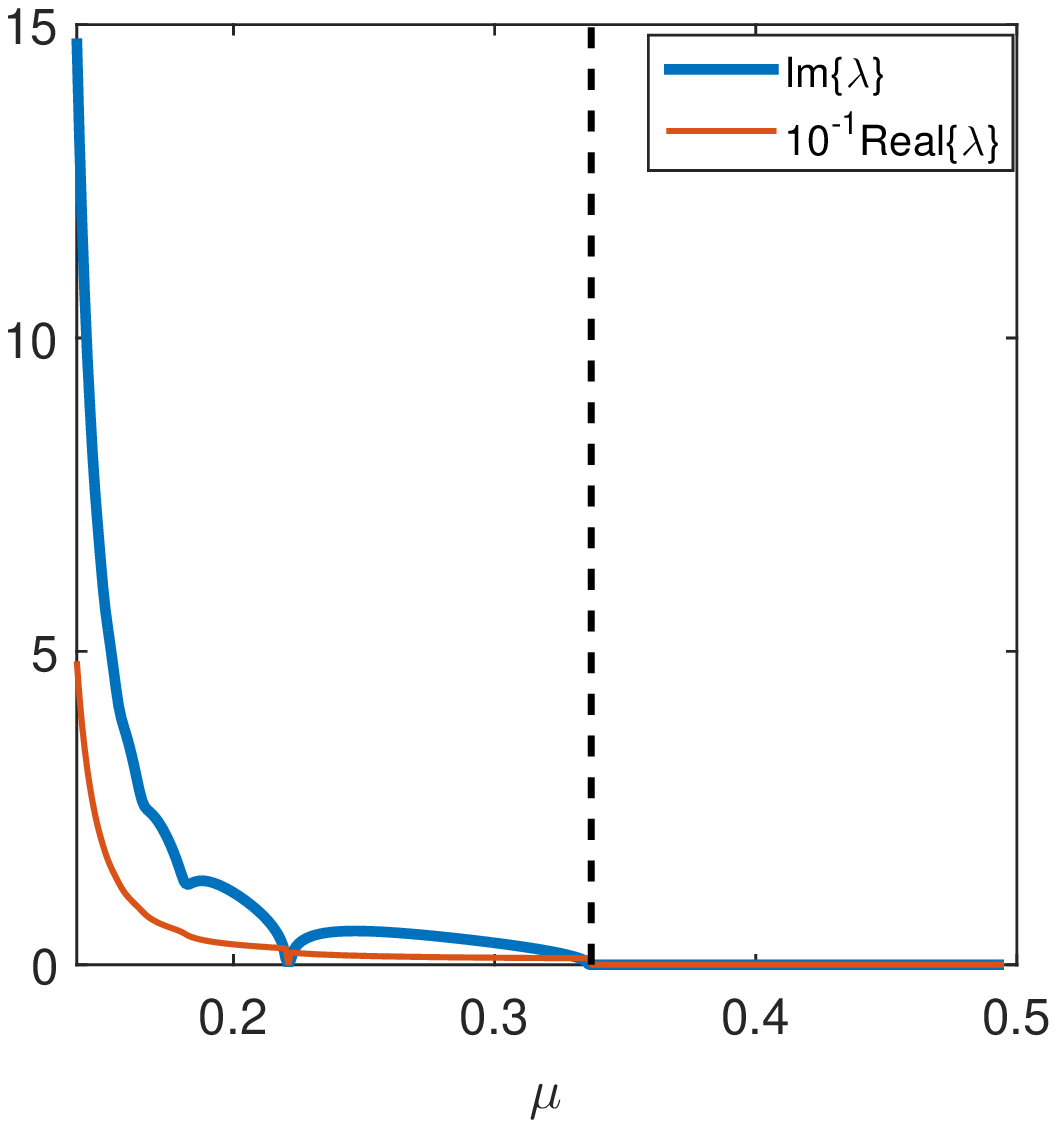}} \subfigure[]{%
\includegraphics[width=2.6in]{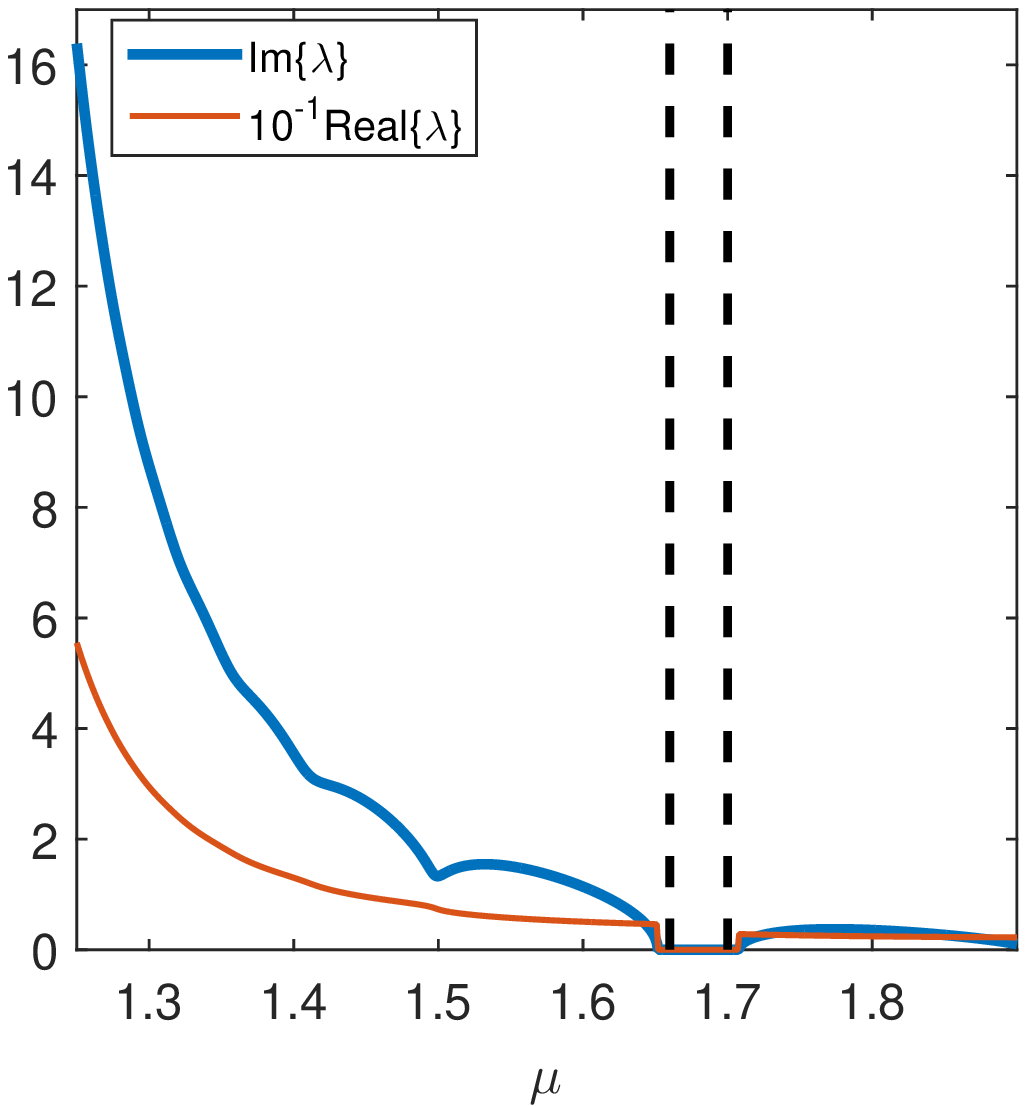}} \subfigure[]{%
\includegraphics[width=2.6in]{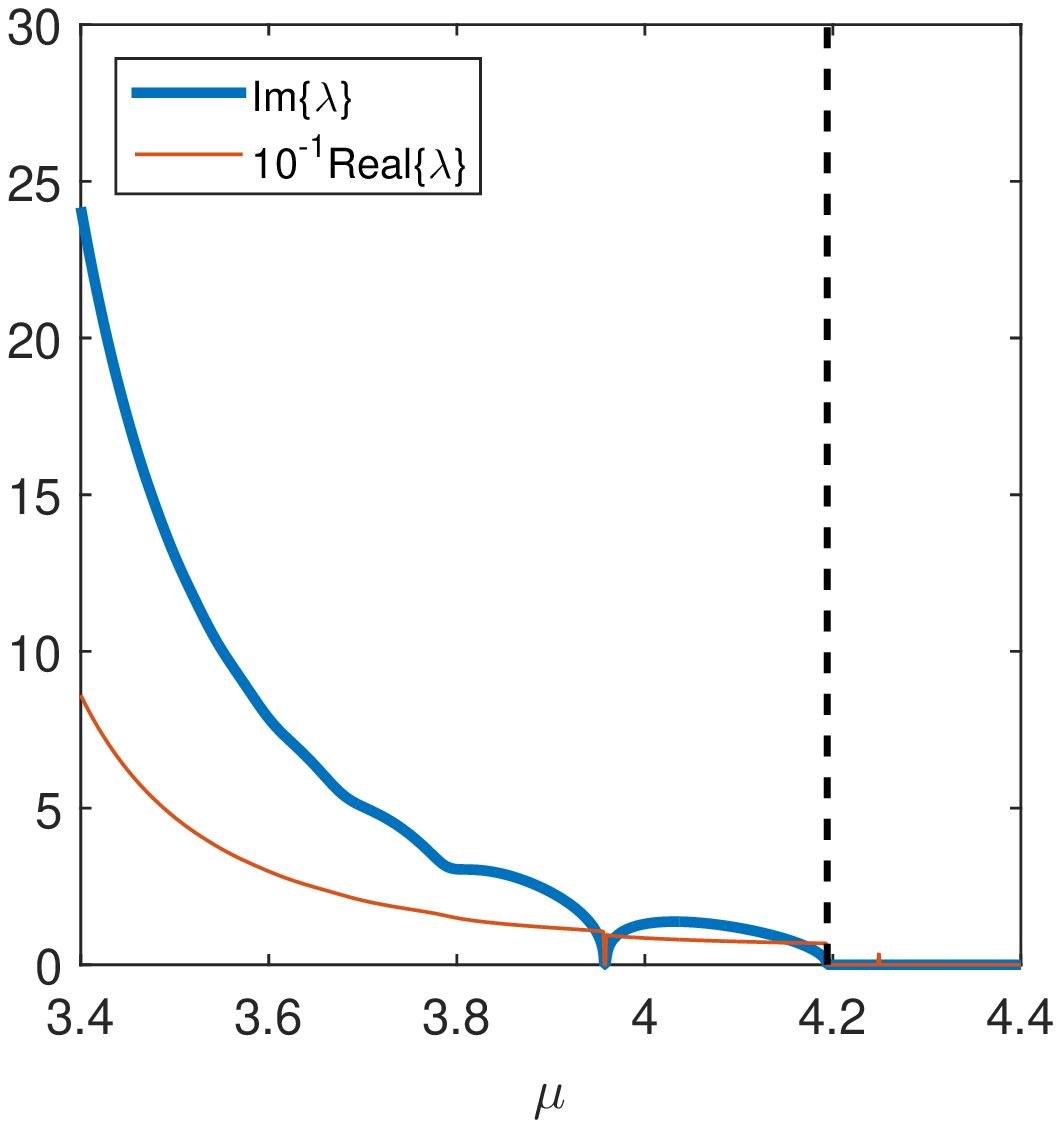}} \subfigure[]{%
\includegraphics[width=2.6in]{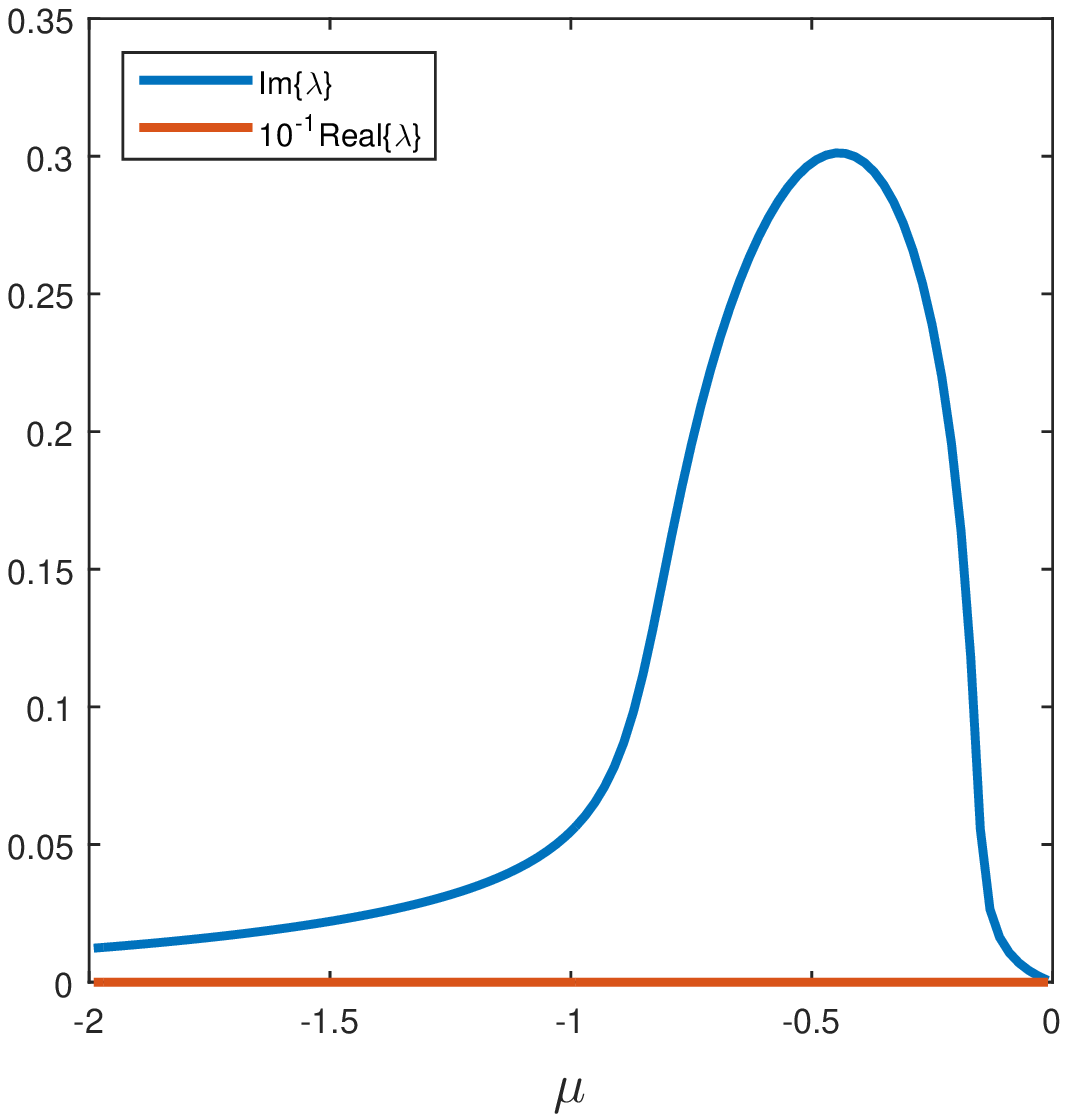}} \subfigure[]{%
\includegraphics[width=2.6in]{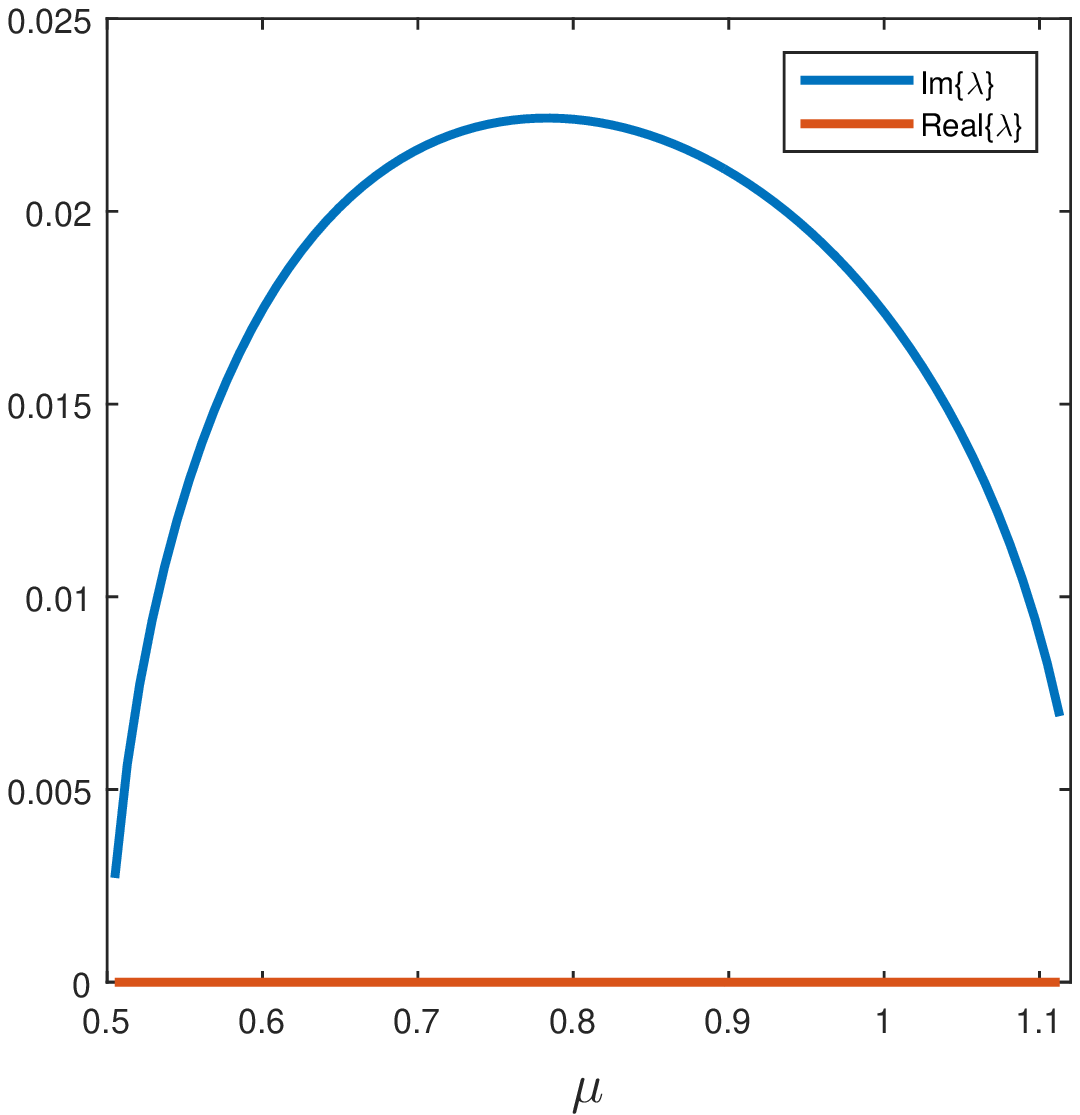}} \subfigure[]{%
\includegraphics[width=2.6in]{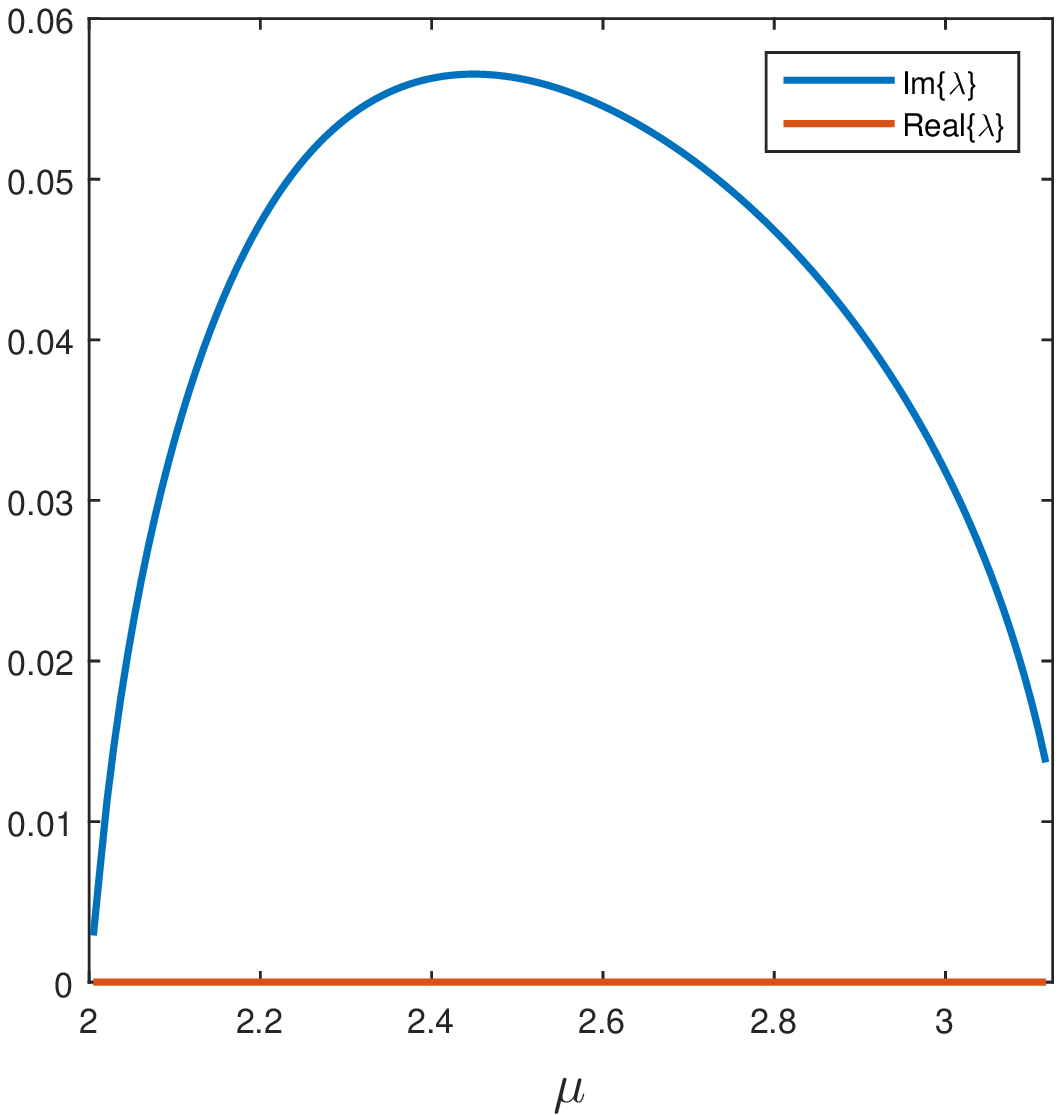}}
\caption{(a-d) Imaginary and real parts of the perturbation eigenfrequency, $%
\protect\lambda $, vs. the chemical potential $\protect\mu >0$, in the
system with $\protect\varepsilon =3$ and $\protect\xi =0.01$. Panels (a),
(b), and (c) display, respectively, the dependences for $\protect\mu >0$ and
$n=0$, $1$, and $2$ (c) in Eq. (\protect\ref{mu_for_pos_epsi}). Panel (d)
displays the dependences for $\protect\mu <0$. Only eigenfrequencies with
the largest instability growth rate, $\mathrm{Im}\{\protect\lambda \}$, are
displayed, the underlying stationary modes being stable at $\mathrm{Im}\{%
\protect\lambda \}=0$. Boundaries between stable and unstable subbands,
which are specified in Table 1, are designated by vertical dashed lines.
Panels (e) and (f) show the same, but in the case of the repulsive
nonlinearity, $\protect\varepsilon =-1$, in the second and third bands,
which correspond to Eq. (\protect\ref{mu_for_neg_epsi}) with $n=1$ (a) and $%
n=2$ (b), respectively [the first band, corresponding to $n=0$, is not shown
here, as it has $\mathrm{Im}(\protect\lambda )\equiv 0$]. Note that $\mathrm{%
Im}\{\protect\lambda \}$ is very small in (e) and (f).}
\label{Fig4}
\end{figure}

In a detailed form, the same results which are summarized in Table 1, are
displayed, in terms of the dependence of the largest instability growth rate
of the perturbation, $\mathrm{Im}\{\lambda \}$, on $\mu $, along with
respective $\mathrm{Re}\{\lambda \}$, in Figs. \ref{Fig4}, where segments
with $\mathrm{Im}\{\lambda \}=0$ represent stable stationary states.
Actually, all complex eigenvalues exist in quartets, $\pm i\mathrm{Im}%
\{\lambda \}\pm \mathrm{Re}\{\lambda \}$, with independent signs $\pm $ in
front of the imaginary and real parts. They reduce to double eigenvalues if
either $\mathrm{Im}\{\lambda \}=0$ or $\mathrm{Re}\{\lambda \}=0$, the
underlying solution being stable in the former case.

We stress that the existence of well-defined stable subbands in the second
and third bands implies that the stationary modes with multi-peak shapes,
see Figs. \ref{Fig1}(b) and (c), may be stable in the present system, while
most previously studied nonlinear systems with self-attractive nonlinearity
admit only the stability of the simplest single-peak modes. On the other
hand, the instability of all the states in each band with the norm exceeding
a certain critical value, $N_{\mathrm{cr}}$, is explained by the fact the
underlying equation (\ref{delta(x)}) on the infinite line, with the ideal
delta-function, gives rise to the critical collapse, because is maintains
the TS family, as mentioned above (TS solutions exist precisely in the case
when the critical collapse occurs \cite{Berge,Fibich,1DTownes}). In
particular, for the first band the numerical result, rescaled back to $%
\varepsilon =1$ [for comparison with Eq. (\ref{delta(x)})], yields
\begin{equation}
N_{\mathrm{cr}}\left( \varepsilon =1\right) =3N_{\mathrm{cr}}\left(
\varepsilon =3\right) \approx 1.94,  \label{cr}
\end{equation}%
i.e., almost exactly twice the above-mentioned critical value for Eq. (\ref%
{delta(x)}), $N_{\mathrm{cr}}=1$. The doubling is explained by the shape of
the mode displayed in Fig. \ref{Fig1}(a): it is seen that approximately half
of the total norm is placed around $\delta (\theta )$, and the other half is
collected around the diametrically opposite position, $\theta =\pi $, the
two peaks being separated by points where the local density vanishes.

According to the analytical result given by Eqs. (\ref{N mu>0}) and (\ref%
{wave_amplitude_first_form}), as well as according to what is seen in Figs. %
\ref{Fig1}(d-f), the $N(\mu )$ dependences in all the three bands satisfy
the VK criterion. However, this criterion does not apply to complex
eigenvalues, being only relevant for purely imaginary ones, with $\mathrm{Re}%
\{\lambda \}=0$ \cite{Berge,Fibich}. This fact explains why only parts of
the three bands carry stable modes, as shown by Table 1 and Figs. \ref{Fig4}%
(a-c).

For the same $\varepsilon >0$, the stability of the single-peak stationary
modes with $\mu <0$, such as the one shown in Fig. \ref{Fig3}(a), is
summarized in Fig. \ref{Fig4}(d). It is seen that, strictly speaking, all
such modes are unstable against perturbations with purely imaginary
eigenfrequencies, in contradiction with the VK criterion, as Fig. \ref{Fig6}
features $dN/d\mu <0$ in interval (\ref{no-VK}). However, the largest value
of $\mathrm{Im}\{\lambda \}$ in this interval is actually very small in
comparison with typical values of instability growth rates in other panels
of Fig. \ref{Fig4}. This fact suggests that the instability of the
stationary solutions may be quite weak in this region, which is corroborated
by direct simulations of the perturbed evolution, see Fig. \ref{Fig6}(b)
below. Actually, the instability of these single-peak modes, which are close
to their counterparts supported by the delta-functional self-attractive
cubic nonlinearity in the infinite domain [cf. solutions (\ref{cosh}) and (%
\ref{sol1})], is a \textquotedblleft remnant" of the instability of modes (%
\ref{sol1}) in the infinite system.

Lastly, results of the linear-stability analysis for the stationary
solutions with $\mu >0$, found in the model with the repulsive nonlinearity,
$\varepsilon <0$, are summarized in Figs. \ref{Fig4}(e,f). These solutions
are completely stable in the first bandgap, which corresponds to $n=0$ in
Eq. (\ref{mu_for_neg_epsi}). This result is very natural, as the ground
state in the model with the repulsive nonlinearity, which populates the
lowest band, must be definitely stable. On the other hand, the excited
states populating the second and third bands [which correspond,
respectively, to $n=1$ and $2$ in Eq. (\ref{mu_for_neg_epsi}], are weakly
unstable, in formal contradiction with the anti-VK criterion. However, this
weak instability does not really destroy the stationary modes from the
second and third bands, see below. 

\subsection{Simulations of the perturbed evolution of stationary modes}

The predictions for the stability and instability of the stationary modes,
produced above by the solution of the BdG equations, have been verified by
comparison with direct simulations of Eq. (\ref{NLS}), in which the input
was taken as the corresponding stationary modes with small random
perturbations added to them. First, all the solutions which were predicted
to be stable, \textit{viz}., those belonging to the stable subbands in Table
1in the case of $\varepsilon >0$ and $\mu <0$, as well as ones belonging to
the first band in the case of $\varepsilon <0$ and $\mu >0$, see Fig. \ref%
{Fig2}(d), were found to be stable in the direct simulations, while the
solutions belonging to unstable regions in Table 1 evolve into chaotic
states (not shown here in detail).

Further, in the case of $\varepsilon <0$ and $\mu >0$, the perturbed
evolution of two- and four-peak modes belonging to the second and third
bands, where the numerical solution of the BdG equations yields very small
values of the instability growth rate in Figs.
\index{Fig4}(f,e), demonstrates small-amplitude oscillations around the
persistent stationary modes (not shown here in detail). In other words,
these modes remain effectively stable ones. 

\begin{figure}[tbp]
\subfigure[]{\includegraphics[width=3.2in]{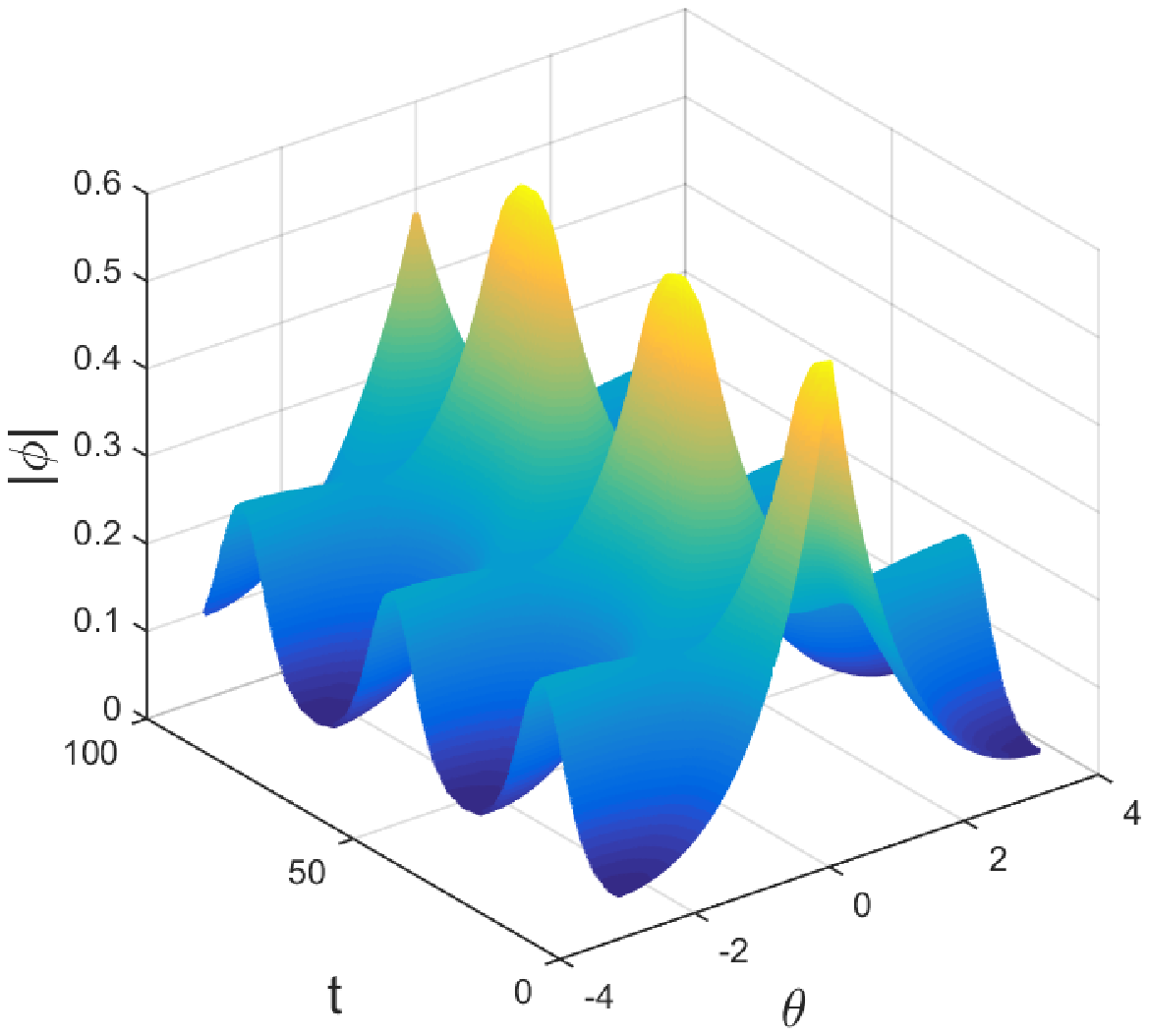}}\subfigure[]{%
\includegraphics[width=3.2in]{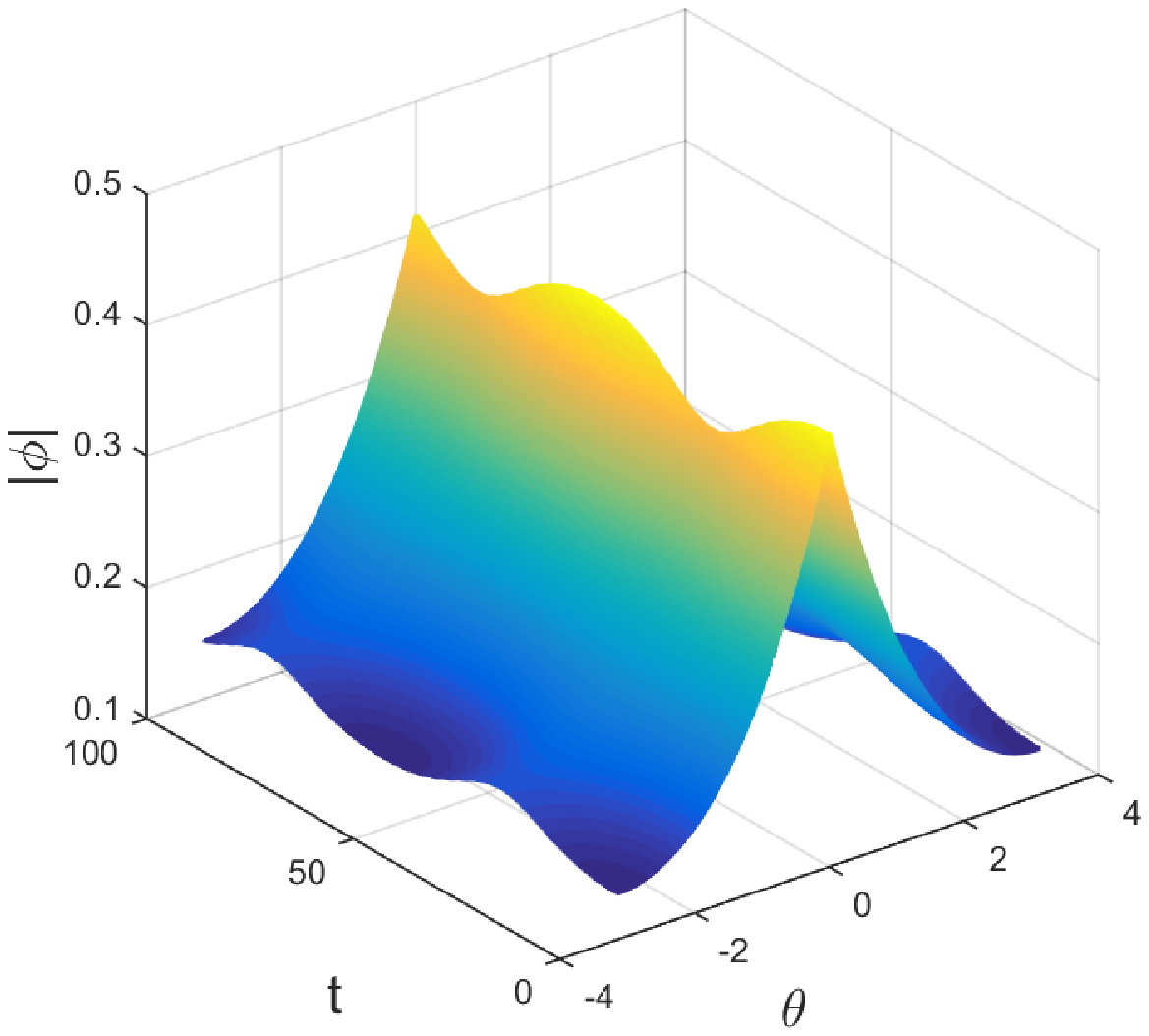}}
\caption{Direct simulations of the evolution of initially perturbed modes
with $\protect\mu <0$ for $\protect\varepsilon =3$ and $\protect\xi =0.01$,
which are predicted to be unstable by the linear-stability analysis, see
Fig. \protect\ref{Fig4}(d). Panels (a) and (b) pertain, respectively, to
original stationary modes with $\protect\mu =-0.5$, $N=0.344$ and $\protect%
\mu =-0.17,N=0.36$, the latter one being close to the state with the largest
norm, whose parameters are given by Eqs. (\protect\ref{Nmax}) and (\protect
\ref{mumax}).}
\label{Fig6}
\end{figure}

Lastly, direct simulations of the perturbed evolution of the stationary
modes which are found, at $\varepsilon >0$, with $\mu <0$ demonstrate that
their instability, predicted by the computation of perturbation eigenvalues
in Fig. \ref{Fig4}(d), leads to spontaneous transformation of the modes into
robustly oscillating \textit{breathers}, as shown in Fig. \ref{Fig6}. The
stationary solution which carries the largest instability growth rate, at $%
\mu =-0.5$ in Fig. \ref{Fig4}(d), develops into a breather with
large-amplitude oscillations [Fig. \ref{Fig6}(a)]. On the other hand, Fig. %
\ref{Fig6}(b) shows that, for $\mu $ close to point (\ref{mumax}), at which
the norm of the stationary model attains the maximum value (\ref{Nmax}), and
the respective instability growth rate in Fig. \ref{Fig4}(d) is quite small,
the resulting breather features small-amplitude oscillations, hence it may
be categorized as a nearly stable state.

\section{Conclusion}

The objective of this work is to generate basic stationary states in the
model which is based on the one-dimensional Schr\"{o}dinger equation on a
ring, with the nonlinearity, of either attractive or repulsive sign, $%
\varepsilon >0$ or $\varepsilon <0$, concentrated at a single point, which
may be represented by an ideal delta-function, or by a regularized narrow
profile. In the case of the ideal delta-function, all the stationary
solutions have been found in the exact analytical form. In particular, the
stationary states with positive chemical potential, $\mu >0$, populate bands
alternating for $\varepsilon >0$ / $\varepsilon <0$, and the stationary
states with $\mu <0$ fill a semi-infinite band, solely for $\varepsilon >0$.
While the exact solutions for $\mu <0$ are qualitatively similar to those
previously reported for the ring with a pair of localized nonlinearities
\cite{HPu}, those for $\mu >0$, which were not found in previous works,
provide a unique insight into exact bandgap states in the nonlinear model.
The stability of the stationary states was investigated by means of
numerical methods (the computation of perturbation eigenfrequencies and
direct simulations of the perturbed evolution) in the three lowest bands for
$\mu >0$, revealing that each band, in the case of $\varepsilon >0$, is
split into stable and unstable subbands, the stability loss occurring with
the increase of the total norm, $N$, at some critical level. Thus,
multi-peak states, populating the higher (second and third) bands, are
partly stable in them. In the case of the repulsive nonlinearity, the
single-peak ground state is completely stable in the first band, while the
two- and four-peak excited states in the two higher bands are weakly
unstable in terms of their eigenvalues, staying virtually stable in direct
simulations. In the case of the attractive self-interaction, the bound
states with $\mu <0$ are subject to the instability which spontaneously
transforms them into robust breathers. The latter instability is weak for
the states with the norm close to the maxim value.

The analysis reported in this work can be extended for a more general
setting, when a narrow stripe carrying a higher-order nonlinearity (e.g.,
quintic) is embedded into a medium with the uniformly distributed
nonlinearity of a different type (e.g., cubic), as suggested by the analysis
performed for the infinite domain in Ref. \cite{Radha}. Another relevant
direction may be the analysis of a model similar to the present one, but
with two components, governed by nonlinearly coupled NLSEs, cf. Ref. \cite%
{Shnir}.

\section*{Acknowledgment}

This work was supported, in part, by the Binational (US-Israel) Science
Foundation through grant No. 2015616.


\begin{thebibliography}{99}
\bibitem{FA} F. Kh. Abdullaev and J. Garnier, Propagation of matter-wave
solitons in periodic and random nonlinear potentials, Phys. Rev. A \textbf{72%
}, 061605(R).

\bibitem{PGK} G. Theocharis, P. Schmelcher, P. G. Kevrekidis, and D. J.
Frantzeskakis, Matter-wave solitons of collisionally inhomogeneous
condensates, Phys. Rev. A \textbf{72}, 033614 (2005).

\bibitem{RMP} Y. V. Kartashov, B. A. Malomed, and L. Torner, Solitons in
nonlinear lattices, Rev. Mod. Phys. \textbf{83}, 247-306 (2011).

\bibitem{GPE} L. P. Pitaevskii and A. Stringari, \textit{Bose-Einstein
Condensation} (Clarendon Press, Oxford, 2003).

\bibitem{Mark} B. A. Malomed and M. Ya. Azbel, Modulational instability of a
wave scattered by a nonlinear center, Phys. Rev. B \textbf{4}7, 10402-10406
(1993).

\bibitem{KA} Y. S. Kivshar and G. P. Agrawal, \textit{Optical Solitons: From
Fibers to Photonic Crystals} (Academic, San Diego, 2003).

\bibitem{Kip} J. Hukriede, D. Runde, and D. Kip, Fabrication and application
of holographic Bragg gratings in lithium niobate channel waveguides, J.
Phys. D \textbf{36}, R1-R16 (2003).

\bibitem{Painting} K. Henderson, C. Ryu, C. MacCormick,and M. G. Boshier,
Experimental demonstration of painting arbitrary and dynamic potentials for
Bose-Einstein condensates, New J. Phys. \textbf{11}, 043030 (2009).

\bibitem{FR1} D. M. Bauer, M. Lettner, C. Vo, G. Rempe and S. D\"{u}rr,
Control of a magnetic Feshbach resonance with laser light, Nature Phys. 5,
339-342 (2009).

\bibitem{FR2} R. Yamazaki, S. Taie, S. Sugawa, and Y. Takahashi, Submicron
spatial modulation of an interatomic interaction in a Bose-Einstein
condensate, Phys. Rev. Lett. \textbf{105}, 050405 (2010).

\bibitem{FR3} L. W. Clark, L.-C. Ha, C.-Y. Xu, and C. Chin, Quantum dynamics
with spatiotemporal control of interactions in a stable Bose-Einstein
condensate, Phys. Rev. Lett. \textbf{115}, 153301 (2015).

\bibitem{Townes} R. Y. Chiao, E. Garmire, and C. H. Townes, Self-trapping of
optical beams, Phys. Rev. Lett. \textbf{13}, 479-482 (1964).

\bibitem{VK} M. Vakhitov and A. Kolokolov, Stationary solutions of the wave
equation in a medium with nonlinearity saturation, Radiophys. Quant.
Electron. \textbf{16}, 783-789 (1973).

\bibitem{Berge} L. Berg\'{e}, Wave collapse in physics: principles and
applications to light and plasma waves, Phys. Rep. \textbf{303}, 259-370
(1998).

\bibitem{Fibich} G. Fibich, \textit{The Nonlinear Schr\"{o}dinger Equation:
Singular Solutions and Optical Collapse} (Springer: Heidelberg, 2015).

\bibitem{Nir} N. Dror and B. A. Malomed, Solitons supported by localized
nonlinearities in periodic media, Phys. Rev. A \textbf{83}, 033828 (2011).

\bibitem{1DTownes} F. Kh. Abdullaev and M. Salerno, Gap-Townes solitons and
localized excitations in low-dimensional Bose-Einstein condensates in
optical lattices, Phys. Rev. A \textbf{72}, 033617 (2005).

\bibitem{gap1} V. A. Brazhnyi and V. V. Konotop, Theory of nonlinear matter
waves in optical lattices, Mod. Phys. Lett. B \textbf{18}, 627-651 (2004).

\bibitem{gap2} O. Morsch and M. Oberthaler, Dynamics of Bose-Einstein
condensates in optical lattices, Rev. Mod. Phys. 78, 179-215 (2006).

\bibitem{gap3} F. Lederer, G. I. Stegeman, D. N. Christodoulides, G.
Assanto, M. Segev, and Y. Silberberg, Discrete solitons in optics, Phys.
Rep. \textbf{463}, 1-126 (2008).

\bibitem{HS} H. Sakaguchi and B. A. Malomed, Solitons in combined linear and
nonlinear lattice potentials, Phys. Rev. A \textbf{81}, 013624 (2010).

\bibitem{HS2} H. Sakaguchi and B. A. Malomed, Matter-wave soliton
interferometer based on a nonlinear splitter, New J. Phys. \textbf{18},
025020 (2016).

\bibitem{BenLi} H. Sakaguchi, B. Li, E. Ya. Sherman, and B. A. Malomed,
Composite solitons in two-dimensional spin-orbit coupled self-attractive
Bose-Einstein condensates in free space, Romanian Rep. Phys. \textbf{70},
502 (2018).

\bibitem{Thaw} T. Mayteevarunyoo, B. A. Malomed, and G. Dong, Spontaneous
symmetry breaking in a nonlinear double-well structure,. Phys. Rev. A
\textbf{78}, 053601 (2008).

\bibitem{Shnir} A. Acus, B. A. Malomed, and Y. Shnir, Spontaneous symmetry
breaking of binary fields in a nonlinear double-well structure, Physica D
\textbf{241}, 987 - 1002 (2012).

\bibitem{BECring-theory} R. Sch\"{u}tzhold, M. Uhlmann, Y. Xu, and U. R.
Fischer, Sweeping from the superfluid to the Mott phase in the Bose-Hubbard
model, Phys. Rev. Lett. \textbf{97}, 200601 (2006).

\bibitem{BECring1} J. A. Sauer, M. D. Barrett, and M. S. Chapman, Storage
ring for neutral atoms, Phys. Rev. Lett. \textbf{87}, 270401 (2001).

\bibitem{BECring2} S. Gupta, K. W. Murch, K. L. Moore, T. P. Purdy, and D.
M. Stamper-Kurn, Bose-Einstein condensation in a circular waveguide, Phys.
Rev. Lett. \textbf{95}, 143201 (2005).

\bibitem{BECring3} A. S. Arnold, C. S. Garvie, and E. Riis, Large magnetic
storage ring for Bose-Einstein condensates, Phys. Rev. A \textbf{73},
041606(R) (2006).

\bibitem{BECring4} O. Morizot, Y. Colombe, V. Lorent, H. Perrin, and B. M.
Garraway, Ring trap for ultracold atoms, Phys. Rev. A \textbf{74}, 023617
(2006).

\bibitem{weak-link1} A. Ramanathan, K. C. Wright, S. R. Muniz, M. Zelan, W.
T. Hill, III, C. J. Lobb, K. Helmerson, W. D. Phillips, and G. K. Campbell,
Superflow in a toroidal Bose-Einstein condensate: An atom circuit with a
tunable weak link, Phys. Rev. Lett. \textbf{106}, 130401 (2011).

\bibitem{BECring6} B. E. Sherlock, M. Gildemeister, E. Owen, E. Nugent, and
C. J. Foot, Time-averaged adiabatic ring potential for ultracold atoms,
Phys. Rev. A \textbf{83}, 043408 (2011); Erratum: Phys. Rev. A \textbf{83},
059904 (2011).

\bibitem{BECring7} S. Moulder, S. Beattie, R. P. Smith, N. Tammuz, and Z.
Hadzibabic, Quantized supercurrent decay in an annular Bose-Einstein
condensate, Phys. Rev. A \textbf{86}, 013629 (2012).

\bibitem{weak-link2} K. C. Wright, R. B. Blakestad, C. J. Lobb, W. D.
Phillips, and G. K. Campbell, Driving phase slips in a superfluid atom
circuit with a rotating weak link, Phys. Rev. Lett. \textbf{110}, 025302
(2013).

\bibitem{weak-link3} S. Eckel, J. G. Lee, F. Jendrzejewski, N. Murray, C. W.
Clark, C. J. Lobb,W. D. Phillips, M. Edwards, and G. K. Campbell, Hysteresis
in a quantized superfluid ``atomtronic" circuit, Nature \textbf{506},
200-203 (2014).

\bibitem{omni1} S. G. Johnson, M. Ibanescu, M. Skorobogatiy, O. Weisberg, T.
D. Engeness, M. Solja\v{c}i\'{c}, S. A. Jacobs, J. D. Joannopoulos, and Y.
Fink, Low-loss asymptotically single-mode propagation in large-core
OmniGuide fibers, Opt. Exp. \textbf{9}, 748-779 (2001).

\bibitem{omni2} M. Ibanescu, S. G. Johnson, M. Solja\v{c}i\'{c}, J. D.
Joannopoulos, Y. Fink, O. Weisberg, T. D. Engeness, S. A. Jacobs, and M.
Skorobogatiy, Analysis of mode structure in hollow dielectric waveguide
fibers, Phys. Rev. E \textbf{67}, 046608 (2003).

\bibitem{omni3} Z. Ruff, D. Shemuly, X. A. Peng, O. Shapira, Z. Wang, and Y.
Fink, Polymer-composite fibers for transmitting high peak power pulses at
1.55 microns, Opt. Exp. \textbf{18}, 15697-15703 (2010).

\bibitem{Bragg1} J. Scheuer and A. Yariv, Coupled-waves approach to the
design and analysis of Bragg and photonic crystal annular resonators, IEEE
J. Quant. Elect. \textbf{39}, 1555-1562 (2003).

\bibitem{Bragg2} J. Scheuer and A. Yariv, Circular photonic-crystal
resonators, Phys. Rev. E \textbf{70}, 036603 (2004).

\bibitem{Bragg3} J. Scheuer, W. M. J. Green, G. A. DeRose, and A. Yariv,
Lasing from a circular Bragg nanocavity with an ultrasmall modal volume,
Appl. Phys. Lett. \textbf{86}, 251101 (2005).

\bibitem{Koby} J. Scheuer and B. Malomed. Annular gap solitons in Kerr media
with circular gratings. Phys. Rev. A \textbf{75}, 063805 (2007).

\bibitem{Hoq} Q. E. Hoq, P. G. Kevrekidis, D. J. Frantzeskakis, and B. A.
Malomed, Ring-shaped solitons in a ``dartboard" photonic lattice, Phys.
Lett. A \textbf{34}1, 145-155 (2005).

\bibitem{Photorefr-radial} X. Wang, Z. Chen, and P. G. Kevrekidis,
Observation of discrete solitons and soliton rotation in optically induced
periodic ring lattices, Phys. Rev. Lett. \textbf{96}, 083904 (2006).

\bibitem{VCSEL1} D. Burak and R. Binder, Cold-cavity vectorial eigenmodes of
VCSEL's, IEEE J. Quant. Elect. \textbf{33}, 1205-1215 (1997).

\bibitem{VCSEL2} A. Valle, Selection and modulation of high-order transverse
modes in vertical-cavity surface-emitting lasers, IEEE J. Quant. Elect.
\textbf{34}, 1924-1932 (1998).

\bibitem{VCSEL} M. Miller, M. Grabherr, R. King, R. Jager, R. Michalzik, and
K. J. Ebeling, Improved output performance of high-power VCSELs, IEEE J.
Sel. Top. Quant. Elect. \textbf{7}, 210-216 (2001)

\bibitem{VCSEL3} X. F. Li, W. Pan, B. Luo, M. Da, and G. Peng, Theoretical
analysis of multi-transverse-mode characteristics of vertical-cavity
surface-emitting lasers, Semicond. Sci. Tech. \textbf{20}, 505-513 (2005).

\bibitem{plasmonic-fiber} J.-Y. Yan, L. Li, and J. Xiao, Ring-like solitons
in plasmonic fiber waveguides composed of metal-dielectric multilayers, Opt.
Exp. \textbf{20}, 1945-1952 (2012).

\bibitem{plasmonic-fiber2} H. Deng Y. Chen, N. C. Panoiu, B. A. Malomed, and
F. Ye, Surface modes in plasmonic Bragg fibers with negative average
permittivity, Opt. Exp. \textbf{26}, 2559-2568 (2018).

\bibitem{Topol-phot} L. Lu, J. D. Joannopoulos, and M. Solja\v{c}i\'{c},
Topological photonics, Nature Photonics \textbf{8}, 821-829 (2014).

\bibitem{protectedBEC} M. Leder, C. Grossert, L. Sitta, M. Genske, A. Rosch,
and M. Weitz, Real-space imaging of a topologically protected edge state
with ultracold atoms in an amplitude-chirped optical lattice, Nature Commun.
\textbf{7}, 13112 (2016).

\bibitem{acoustics} Z. Yang, F. Gao, X. Shi, X. Lin, Z. Gao. Y. Chong, and
B. Zhang, Topological acoustics, Phys. Rev. Lett. \textbf{114}, 114301
(2015).

\bibitem{PhotInsulator1} A. B. Khanikaev, S. H. Mousavi, W.-K. Tse, M.
Kargarian, A, H. Mac-Donald, and G. Shvets, Photonic topological insulators,
Nature Mater. \textbf{12}, 233-239 (2012).

\bibitem{Segev1} M. C. Rechtsman, J. M. Zeuner, Y. Plotnik, Y. Lumer, D.
Podolsky, F. Deisow, S. Nolte, M. Segev, and A. Szameit, Photonic Floquet
topological insulators, Nature \textbf{496}, 196-200 (2013).

\bibitem{PhotInsulator2} W.-J. Chen, S.-J. Jiang, X.-D. Chen, B. Zhu, L.
Zhou, J.-W. Dong, and C. T. Chan, Experimental realization of photonic
topological insulator in a uniaxial waveguide, Nature Commun. \textbf{5},
5782 (2014).

\bibitem{PhotInsulator3} A. V. Nailov, D. D. Solnyshkov, and G. Malpuech,
Polariton $Z$ topological insulator, Phys. Rev. Lett. \textbf{114}, 116401
(2015).

\bibitem{Segev2} M. A. Bandres, M. C. Rechtsman, and M. Segev, Topological
photonic quasicrystals: Fractal topological spectrum and protected
transport, Phys. Rev. X \textbf{6}, 011016 (2016).

\bibitem{PhotInsulator4} Y. V. Kartashov and D. V. Skryabin, Modulational
instability and solitary waves in polariton topological insulators, Optica
\textbf{3}, 1228-1236 (2016).

\bibitem{PhotInsulator5} Y. V. Kartashov and D. V. Skryabin, Bistable
topological insulator with excitons-polaritons, Phys. Rev. Lett. \textbf{119}%
, 253904 (2017).

\bibitem{PhotInsulator7} D. R. Gulevich, D. Yudin, D. V. Skryabin, I. V.
Iorsh, and I. A. Shelykh, Exploring nonlinear topological states of matter
with exciton-polaritons: Edge solitons in kagome lattice, Sci. Rep. \textbf{7%
}, 1780 (2017).

\bibitem{PhotInsulator6} C. Li, F. Ye, X. Chen, Y. V. Kartashov, A.
Ferrando, L. Torner, and D. V. Skryabin, Lieb polariton topological
insulators, Phys. Rev. B \textbf{97}, 081103(R) (2018).

\bibitem{Segev3} G. Harari, M. A. Bandres, Y. Lumer, M. C. Rechtsman, Y. D.
Chong, M. Khajavikhan, D. N. Christodoulides, and M. Segev, Topological
insulator laser: Theory, Science \textbf{359}, 4003 (2018).

\bibitem{Segev4} M. A. Bandres, S. Wittek, G. Harari, M. Parto, \ J. Ren, M.
Segev, D. N. Christodoulides, and M. Khajavikhan, Topological insulator
laser: Experiment, Science \textbf{359}, eaar4005 (2018).

\bibitem{rotary} Y. V. Kartashov, V. A. Vysloukh, and L. Torner, \ Rotary
solitons in Bessel optical lattices, Phys. Rev. Lett. \textbf{93}, 093904
(2004).

\bibitem{rotary2} Y. V. Kartashov, A. A. Egorov, V. A. Vysloukh, and L.
Torner, Rotary dipole-mode solitons in Bessel optical lattices, J. Opt. B:
Quant. Semicl. Opt. \textbf{6}, 444-447 (2004).

\bibitem{Bakhtiyor} B. Baizakov, B. A. Malomed, and M. Salerno, Matter-wave
solitons in radially periodic potentials, Phys. Rev. E \textbf{74}, 066615
(2006).

\bibitem{Kuzmiak} A. V. Yulin, Yu. V. Bludov, V. V. Konotop, V. Kuzmiak, and
M. Salerno, Superfluidity of Bose-Einstein condensates in toroidal traps
with nonlinear lattices, Phys. Rev. A \textbf{84}, 063638 (2011).

\bibitem{HPu} X.-F. Zhou, S.-L. Zhang, Z.-W. Zhou, B. A. Malomed, and H. Pu,
Bose-Einstein condensation on a ring-shaped trap with nonlinear double-well
potential, Phys. Rev. A \textbf{85}, 023603 (2012).

\bibitem{Yang} J. Yang, \textit{Nonlinear Waves in Integrable and
Nonintegrable Systems} (SIAM:\ Philadelphia, 2010).

\bibitem{Radha} H. Fabrelli, J. B. Sudharsan, R. Radha, A. Gammal, and B. A.
Malomed, Solitons under spatially localized cubic-quintic-septimal
nonlinearities, J. Optics \textbf{19}, 075501 (2017).
\end{thebibliography}
\end{document}